\documentclass{article}

    \PassOptionsToPackage{numbers, compress}{natbib}
\usepackage[preprint]{neurips_2026}

\usepackage[utf8]{inputenc} % allow utf-8 input
\usepackage[T1]{fontenc}    % use 8-bit T1 fonts
\usepackage{hyperref}       % hyperlinks
\usepackage{url}            % simple URL typesetting
\usepackage{booktabs}       % professional-quality tables
\usepackage{amsfonts}       % blackboard math symbols
\usepackage{nicefrac}       % compact symbols for 1/2, etc.
\usepackage{microtype}      % microtypography
\usepackage{xcolor}         % colors

\usepackage{wrapfig}
\usepackage{graphicx}
\usepackage{mathtools}
\usepackage{amsmath, amssymb, amsthm}
\usepackage{bm}
\usepackage{enumitem}
\usepackage{multirow}
\usepackage{makecell}
\usepackage{siunitx}
\usepackage{adjustbox}
\usepackage{tabularx}
\usepackage{diagbox}
\usepackage[most]{tcolorbox}
\usepackage{anyfontsize}
\usepackage{hyperref}       % hyperlinks
\usepackage{url}            % simple URL typesetting
\usepackage{booktabs}       % professional-quality tables
\usepackage{amsfonts}       % blackboard math symbols
\usepackage{nicefrac}       % compact symbols for 1/2, etc.
\usepackage{microtype}      % microtypography
\usepackage{xcolor}         % colors
\usepackage{natbib, mathtools}
\usepackage{multirow}
\usepackage{graphicx}
\usepackage[table]{xcolor}
\usepackage{array}
\usepackage{siunitx}
\usepackage{ragged2e}
\usepackage{listings}

\theoremstyle{plain}

\def\eqref#1{equation~\ref{#1}}
\title{GGBound: A Genome-Grounded Agent for Microbial Life-Boundary Prediction}

\author{
Hanbo Huang\textsuperscript{1~*}
\quad
Xuan Gong\textsuperscript{1~*}
\quad
Jing Wang\textsuperscript{1~*}
\quad
Lei Bai\textsuperscript{2}
\\
\textbf{Xiang Xiao}\textsuperscript{1}
\quad
\textbf{Weishu Zhao}\textsuperscript{1}
\quad
\textbf{Shiyu Liang}\textsuperscript{1~$\dagger$}
\\
\textsuperscript{1}Shanghai Jiao Tong University\quad\textsuperscript{2}Shanghai Artificial Intelligence Laboratory\\
{\tt\small \{hhuang417, gongxuan0610, lsy18602808513\}@sjtu.edu.cn}
}

\begin{document}

\maketitle

\def\thefootnote{*}\footnotetext{Equal contribution.}
\def\thefootnote{$\dagger$}\footnotetext{Corresponding author.}

\begin{abstract}

Characterizing the physiological life boundaries of microbial strains, including viable temperature, pH, salinity, substrate utilization, and morphology, is central to biotechnology and ecology, yet traditionally requires exhaustive \emph{in vitro} screening. Existing computational approaches either treat physiological traits as isolated supervised targets or repurpose biological foundation models as static encoders, leaving the genotype-to-physiology gap largely unbridged. We formulate \emph{microbial life-boundary prediction} as a unified genome-to-physiology task and address it with a genome-conditioned, tool-augmented LLM agent. To support this task, we curate a strain-centric benchmark from IJSEM, NCBI, and BacDive covering $1{,}525$ strains and $6{,}448$ instances across viability intervals, environmental optima, substrate utilization, categorical traits, and morphology. Architecturally, the agent injects frozen LucaOne genome embeddings into a Qwen backbone via lightweight token fusion, and reasons over a similarity-based RAG module and a Genome-scale Metabolic Model (GEM) perturbation tool. We optimize the agent through a three-stage pipeline of gene--text alignment, agentic SFT on distilled trajectories, and GRPO with a novel \emph{counterfactual gene-grounding} reward that reinforces the policy only when the authentic genome embedding causally improves correct-token generation relative to a zero-gene ablation. The resulting $4$B-parameter agent matches or surpasses substantially larger frontier LLMs, with ablations confirming that genome-token fusion, dynamic tool use, and the counterfactual reward each yield distinct, significant gains.
\end{abstract}

\section{Introduction}

Characterizing the physiological \emph{life boundaries} of microbial strains, such as temperature, pH, and substrate tolerances, is fundamental to biotechnology and ecology~\citep{rothschild2001life, harrison2013limits, merino2019living}. However, determining these limits traditionally requires exhaustive \textit{in vitro} screening~\citep{nichols2010use, lagier2018culturing}. Despite the abundance of genomic data~\citep{sayers2021databasencbi}, mapping genotypes to comprehensive physiological envelopes remains a formidable predictive challenge, leaving the genotype-phenotype gap bridged largely by resource-intensive manual experimentation.

Existing predictive methodologies offer incomplete solutions: conventional approaches treat physiological traits as isolated targets using hand-engineered features~\citep{weimann2016genomes, aun2018k, karlsen2023genotype, koblitz2025predicting, drouin2019interpretable}, while biological foundation models serve primarily as static encoders for narrow tasks~\citep{rives2021biological, lin2023evolutionary, ji2021dnabert, zhou2023dnabert, dalla2025nucleotide, nguyen2023hyenadna, nguyen2024sequence, brixi2026genome, he2025generalized}. To bridge this gap, we formulate \emph{microbial life-boundary prediction} as a unified genome-to-physiology paradigm solved via a genome-conditioned, tool-augmented Large Language Model (LLM) agent. Given a strain genome $g_i$ and a physiological query $q_i$, our agent interleaves semantic reasoning with biological tool execution to generate an observation trajectory $O_i=\{o_{i,1},\ldots,o_{i,T}\}$, jointly predicting $\hat{y}_i=f_\theta(g_i,q_i,O_i)$. This predictive target comprehensively encompasses growth ranges, optimal conditions, substrate utilization, and morphological traits.

% Figure~\ref{fig:GGboundOverview} provides an overview of the proposed framework.
% We instantiate this framework through three core contributions. First, we construct a strain-centric benchmark via an automated LLM extraction pipeline over IJSEM, NCBI, and BacDive records, yielding a rigorous dataset covering $1{,}525$ unique strains and $6{,}448$ task-specific instances. Second, we design a genome-grounded agent that injects LucaOne genome embeddings into a Qwen backbone via token fusion and equips the model with dynamic biological tools, including a similarity-based Retrieval-Augmented Generation (RAG) module and a Genome-scale Metabolic Model (GEM) perturbation tool for simulating growth under targeted nutrient variations. Third, we optimize the agent via a three-stage pipeline consisting of gene--text instruction tuning, agentic supervised fine-tuning, and Group Relative Policy Optimization (GRPO)~\citep{shao2024deepseekmath}. Crucially, our GRPO stage employs a counterfactual gene-grounding reward that penalizes reliance on language priors and reinforces the policy only when the authentic genome embedding causally improves accurate token generation.

\begin{figure}[t]
    \centering
    \includegraphics[width=0.95\linewidth]{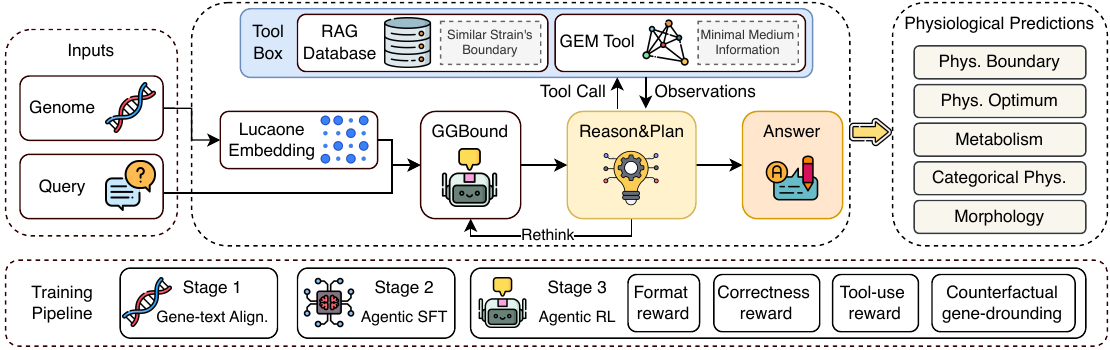}
    \vspace{-0.1cm}
    \caption{Overview of GGBound. The agent conditions on LucaOne genome embeddings, integrates RAG and GEM tool observations, and predicts microbial life boundaries through a three-stage training pipeline with counterfactual gene-grounding.}
    \label{fig:GGboundOverview}
    \vspace{-0.4cm}
\end{figure}

Figure~\ref{fig:GGboundOverview} provides an overview of the proposed framework.
We instantiate this framework through three core contributions. First, we construct a strain-centric benchmark via an automated LLM extraction pipeline over IJSEM, NCBI, and BacDive records, covering $1{,}525$ unique strains and $6{,}448$ task-specific instances. Second, we design a genome-grounded agent that injects LucaOne genome embeddings into a Qwen backbone via token fusion and equips it with dynamic biological tools, including a similarity-based Retrieval-Augmented Generation (RAG) module and a Genome-scale Metabolic Model (GEM) perturbation tool for targeted nutrient perturbations. Third, we optimize the agent through gene--text instruction tuning, agentic supervised fine-tuning, and Group Relative Policy Optimization (GRPO)~\citep{shao2024deepseekmath}. Crucially, the GRPO stage employs a counterfactual gene-grounding reward that penalizes reliance on language priors and reinforces the policy only when the authentic genome embedding causally improves accurate token generation.

Empirically, our $4$B-parameter agent matches or outperforms substantially larger general-purpose LLMs (e.g., DeepSeek-V3.2, DeepSeek-R1, GLM-4.7, Kimi-K2) on life-boundary prediction. Ablations confirm that our CLS-based encoding, dynamic tool-use formulation, and counterfactual reward each yield distinct, significant performance gains. In summary, our primary contributions are:
\begin{itemize}[leftmargin=*, itemsep=1pt, parsep=0pt, topsep=0pt]
    \item \textbf{Task Formulation \& Benchmark Creation:} We formulate \emph{microbial life-boundary prediction} as a unified genome-to-physiology task. To support this, we construct and release a comprehensive, strain-centric benchmark curated from IJSEM, NCBI, and BacDive, which jointly evaluates viability intervals, environmental optima, substrate utilization, categorical traits, and morphology.
    
    \item \textbf{Genome-Conditioned Agent Architecture:} We design a tool-augmented agent that seamlessly injects LucaOne genome embeddings into a Qwen LLM backbone. The agent reasons over a similarity-based RAG module and a GEM-based perturbation tool, effectively synthesizing evidence across database lookups, embedding spaces, and metabolic simulations.
    
    \item \textbf{Counterfactual RL Pipeline \& Empirical Performance:} We propose a three-stage training paradigm: gene--text alignment, agentic SFT on distilled trajectories, and GRPO. Crucially, we introduce a \emph{counterfactual gene-grounding} reward that explicitly enforces causal dependence on the genomic input. This pipeline yields a $4$B-parameter agent that achieves performance competitive with substantially larger frontier LLMs.
\end{itemize}

\section{Related Work}

\textbf{Biological Sequence-to-Phenotype Modeling.}
Mapping biological sequences to phenotypic traits is a central challenge in computational biology. Protein structure prediction has demonstrated that amino-acid sequences contain sufficient signal for accurate three-dimensional structure prediction~\citep{jumper2021highly,baek2021accurate,lin2023evolutionary}, while protein language models show that large-scale sequence pretraining captures transferable sequence--structure and sequence--function information~\citep{rives2021biological,elnaggar2021prottrans}. In microbial genomics, phenotype prediction has been studied using k-mer features, protein-family inventories, phyletic patterns, and supervised learning methods, including tools such as Traitar and PhenotypeSeeker as well as recent BacDive-based models for bacterial physiological traits~\citep{weimann2016genomes,aun2018k,karlsen2023genotype,koblitz2025predicting}. Prior microbial phenotype predictors usually optimize trait-specific supervised models or metabolic simulations, whereas our system evaluates whether a genome-conditioned agent can combine representation similarity, database evidence, and GEM-derived observations across heterogeneous physiological targets.

\textbf{Biological Foundation Models and Genome-Grounded Agents.}
Biological foundation models have enabled scalable representation learning across proteins, nucleotides, and genomes. Protein models such as ESM and ProtTrans learn transferable representations from large-scale amino-acid corpora~\citep{lin2023evolutionary,rives2021biological,elnaggar2021prottrans}, while nucleotide and genome-scale models such as DNABERT, DNABERT-2, Nucleotide Transformer, HyenaDNA, GenSLM, Evo, and Evo~2 extend sequence modeling to regulatory, mutational, evolutionary, and long-context genome-scale prediction~\citep{ji2021dnabert,zhou2023dnabert,dalla2025nucleotide,nguyen2023hyenadna,zvyagin2023genslms,nguyen2024sequence,brixi2026genome}. LucaOne further unifies nucleic-acid and protein modeling within a single biological foundation model~\citep{he2025generalized}. However, these models are commonly used as static encoders, zero-shot scorers, or backbones for task-specific predictors. Our work instead treats dense genome embeddings as continuous, non-textual conditioning inputs for a tool-augmented LLM agent. Building on multimodal LLM alignment and tool-use agents~\citep{li2023blip,liu2023visual,yao2022react,schick2023toolformer,wang2025biobridge}, we integrate genome-token fusion, biological tool use, and GRPO-based optimization~\citep{shao2024deepseekmath}, training the agent to gather biological evidence and align its predictions with the underlying genomic context.
\section{Motivation}

% We introduce genome-conditioned microbial life-boundary prediction, a task that infers physiological viability regimes from genomic representations and tool-derived biological observations, thereby enabling computational studies of microbial life limits and adaptation in extreme environments.
We study strain-level prediction of microbial physiological regimes, including growth-condition ranges, optimal conditions, substrate utilization, and morphology, using genome-derived embeddings and external biological evidence.

\textbf{Microbial life-boundary prediction is an underexplored genome-to-physiology problem.}
Determining the environmental regimes under which microbial strains survive and grow is fundamental to cultivation, experimental design, ecological characterization, and biological adaptation~\citep{rothschild2001life, harrison2013limits, merino2019living}. 
Such characterization often requires extensive wet-lab screening over combinatorial spaces of temperature, pH, salinity, pressure, nutrient availability, and other factors~\citep{nichols2010use, lagier2018culturing}. 
Although genome sequencing has made strain-level molecular information increasingly accessible, existing genotype-to-phenotype studies typically formulate microbial traits as separate categorical or scalar prediction problems~\citep{drouin2019interpretable, karlsen2023genotype}. 
We instead formulate microbial life-boundary prediction as a broader genome-to-physiology problem, where boundary-level viability regimes serve as the central prediction target and organize strain-level physiological inference.

\textbf{Tool observations provide a general interface for integrating biological evidence.}
Inferring physiological boundaries from genomic information alone is challenging, as microbial viability is governed by coupled constraints from evolutionary relatedness, cultivation history, environmental context, phenotype annotations, and metabolic feasibility~\citep{sayers2021databasencbi, schober2025bacdive, thiele2010protocol, heirendt2019creation}. 
Rather than defining source-specific modules for each type of evidence, we represent external biological information through a unified sequence of tool observations, following recent tool-augmented agent formulations~\citep{yao2022react, schick2023toolformer}. 
Given a genome representation $g_i$ and a task query $q_i$, the agent interacts with biological tools and obtains an observation trajectory $O_i = \{o_{i,1}, \ldots, o_{i,T}\}$, where each observation provides task-relevant evidence about the strain. 
The prediction objective is defined as
\begin{align*}
    \hat{y}_i = f_{\theta}(g_i, q_i, O_i).
\end{align*}
This formulation provides a unified basis for incorporating heterogeneous biological evidence into strain-level physiological prediction across diverse target spaces.

\textbf{Genome-conditioned life-boundary agents can accelerate experimental discovery.}
The goal of life-boundary prediction is to generate biologically plausible and experimentally actionable hypotheses for poorly characterized strains, rather than merely reproduce annotations. 
By prioritizing feasible cultivation conditions and excluding implausible regimes, accurate boundary predictions can reduce the wet-lab search space for microbial characterization~\citep{nichols2010use, lagier2018culturing}. 
More broadly, linking genomic information to viability regimes provides a computational framework for studying microbial adaptation in extreme environments, including deep-sea, high-pressure, hypersaline, and acidic habitats~\citep{rothschild2001life, harrison2013limits, merino2019living}. 
These considerations further motivate our strain-centric genome-physiology dataset and tool-augmented agentic framework for microbial life-boundary prediction.

\section{Data Collection and Benchmark Construction}
Building a genome-conditioned life-boundary agent requires aligned strain-level genome-physiology data. 
We therefore integrate distributed literature, genomic, and metabolic resources into a unified corpus, from which we construct standardized benchmarks and agent training data. (Appendix~\ref{app:dataset_benchmark}.)

\subsection{Genome-Physiology Corpus Construction and Modality Alignment}

\textbf{Construction of a strain-centric genome-physiology corpus.}
We constructed a comprehensive microbial physiology corpus by integrating heterogeneous biological resources, including IJSEM~\cite{ijsem} articles, NCBI~\cite{sayers2021databasencbi} genome and protein records, and BacDive~\cite{schober2025bacdive} phenotype annotations. Specifically, we first extracted microbial strains and their associated physiological traits from IJSEM publications, followed by systematic data cleaning and normalization. Using the curated strain identities, we retrieved corresponding genomic and proteomic profiles from NCBI, alongside phenotype annotations from BacDive. The resulting dataset consolidates heterogeneous multimodal sources into unified strain-level records. Each record encapsulates strain identity, genomic and proteomic information, morphology, metabolism, growth and cultivation conditions, and environmental context, thereby supporting strain-centric analysis of genome-physiology relationships.

\textbf{Raw strain data collection.}
To acquire raw strain-level physiological data, we developed a four-stage LLM-based extraction pipeline powered by DeepSeek-V3.2~\cite{liu2025deepseek32} and applied it to 18,498 IJSEM articles published prior to Jan. 16, 2026. The pipeline sequentially identifies strain-relevant paragraphs and tables, isolates the primary strains under study, extracts their associated physiological and metabolic traits, and applies an LLM-based quality scoring procedure to filter incomplete or unreliable records. After filtering and normalization, the IJSEM-derived corpus yielded 88,927 high-quality microbial strain records covering morphology, metabolic properties, cultivation protocols, and isolation environments. To assess extraction quality, domain experts manually evaluated a subset of the resulting records, confirming high fidelity and structural consistency.

\textbf{Multi-modal data alignment.} 
We aligned the literature-derived records with external genomic, phenotypic, and metabolic databases. For strains extracted from IJSEM, we retrieved corresponding genome and protein records from NCBI, identifying 43,649 matched instances and 18,964 unique strains post-filtering. Utilizing scientific nomenclature and standardized strain identifiers, these records were further mapped to BacDive annotations, yielding 17,796 successfully matched strains. Concurrently, gene sequences were encoded using LucaOne~\citep{he2025generalized} to generate gene-level embeddings, which were subsequently aggregated into strain-level representations. 
Given that the classification (CLS) token encapsulates sufficient global semantic information for downstream tasks such as morphological classification~\citep{he2025generalized}, we directly utilize the CLS token from the genome-level LucaOne embedding as a compact strain representation, yielding a $1 \times 2560$ dimensional vector that serves as the genetic modality input for our agent. To empirically verify that this compact representation preserves signal for physiological-boundary prediction, we conduct ablations against mean pooling, gene-set pooling, protein-family inventories, and k-mer baselines.

\textbf{Trait harmonization.}
Physiological traits are often reported with inconsistent terminology, measurement units, and levels of granularity across the literature. We therefore systematically standardized these descriptions. Continuous variables (e.g., temperature, pH, and salinity) were reformatted into unified scalar or interval representations. Meanwhile, categorical attributes (e.g., cellular morphology, motility, and oxygen tolerance) were rigorously mapped to controlled vocabularies developed in consultation with biological domain experts.

\subsection{Benchmark and Agent Training Data Construction}

\begin{figure}[t]
    \centering
    \includegraphics[width=\linewidth]{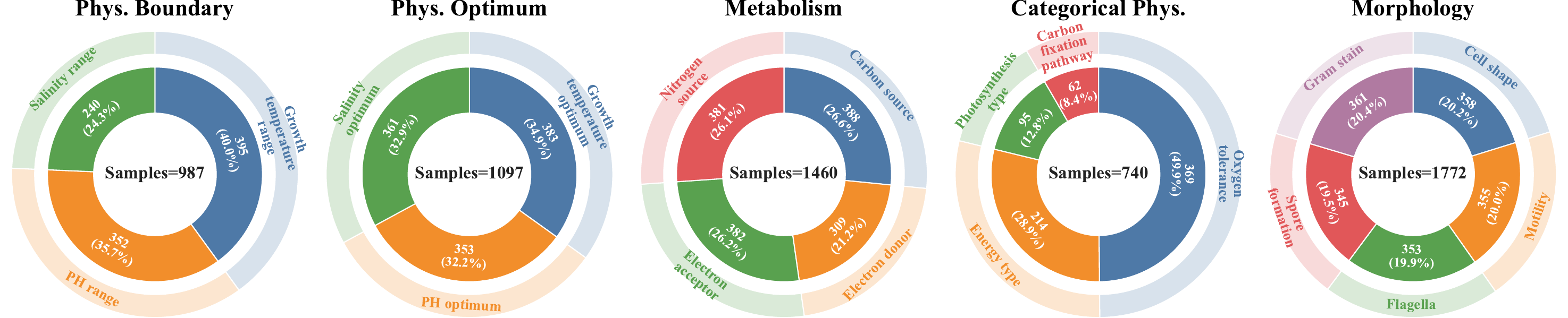}
    \caption{Benchmark composition. The benchmark contains $6{,}448$ strain-level instances across five physiological prediction groups. Bars show task-level counts and proportions.}
    \label{fig:benchfig}
    \vspace{-0.3cm}
\end{figure}

% \textbf{Benchmark construction.}
% To evaluate genome-conditioned microbial physiology prediction, we constructed a benchmark targeting five predictive capabilities: (1) physiological boundary prediction, (2) optimal physiological condition prediction, (3) substrate utilization and metabolic capability prediction, (4) categorical physiological trait prediction, and (5) cellular morphology prediction. Covering 1,525 unique strains, the benchmark comprises a total of 6,448 task-specific instances, each pairing a genome-derived representation with a target attribute. During evaluation, true strain names are fully anonymized to prevent models from exploiting memorized taxonomic knowledge, ensuring predictions rely on genomic features and tool utilization. All benchmark samples are strictly disjoint from downstream agent training data.
\paragraph{Benchmark construction.}
As summarized in Figure~\ref{fig:benchfig}, we construct a strain-centric benchmark covering five predictive capabilities: (1) physiological boundary prediction, (2) optimal-condition prediction, (3) substrate utilization and metabolic capability prediction, (4) categorical physiological trait prediction, and (5) cellular morphology prediction. The benchmark spans $1{,}525$ strains and $6{,}448$ instances, including $987$ physiological-boundary, $1{,}097$ optimal-condition, $1{,}460$ metabolic, $740$ categorical-physiology, and $1{,}772$ morphology instances. Each instance pairs a genome-derived representation with a target attribute. During evaluation, strain names are anonymized to prevent exploitation of memorized taxonomic knowledge, requiring predictions to rely on genomic features and tool utilization. All benchmark samples are disjoint from downstream agent training data.

\textbf{Data construction for modality fusion.}
LucaOne encodes genomic sequences into continuous gene-level embeddings, which are not directly aligned with the natural-language representation space of LLMs~\cite{yang2025qwen3}. We therefore constructed genome-text modality-fusion data to bridge genomic representations and textual physiological descriptions. Specifically, we reorganized IJSEM-derived records into trait-specific question--answer pairs, where each example conditions on a strain identifier and queries a specific physiological trait. In addition to single-trait QA pairs, we employed Qwen3.5-27B~\cite{qwen35} to generate longer-form, multi-trait QA pairs that synthesize coherent phenotypic profiles. To prevent data leakage, the fields used for fusion are disjoint from the data utilized in agent training.

\textbf{Data construction for agent training.}
The agent-training phase consists of supervised fine-tuning (SFT) followed by reinforcement learning (RL). We constructed the training data primarily from BacDive-derived records, leveraging their structured format and high-quality phenotype annotations. We first filtered BacDive entries by standardized strain identifiers and retained 10,000 exactly matched strain records. Similar to benchmark construction, we anonymized strain identities by replacing true strain names with anonymous identifiers and genome-derived representations. This prevents the model from relying on species-name memorization and requires the agent to infer target traits from genome-conditioned features and tool observations. From these records, we randomly selected 8,000 unique strains for SFT data construction and 2,000 unique strains for the RL stage.

\textbf{Trajectory distillation via teacher rollouts.}
To improve the agent's tool-use capabilities, we performed trajectory distillation using the 8,000 SFT-stage samples. For each sample, we first queried a Qwen3.5-27B teacher agent to generate four candidate trajectories. During generation, the teacher agent was allowed to invoke external tools when necessary to gather supporting evidence, including BacDive records from similar strains and minimal-medium evidence for assessing viability under specified growth conditions. Each trajectory consists of a sequence of tool invocations, the corresponding tool observations, intermediate reasoning steps, and a final JSON-formatted prediction. We then evaluated and selected the best trajectory for each sample according to prediction correctness and consistency with the retrieved evidence. 
% Trajectories that were incorrect or insufficiently grounded in tool observations were passed to a subsequent refinement stage.
We separate answer-supervised correction from evidence-grounded trajectory imitation. Repaired trajectories are not used to train tool-use rationales unless the corrected answer is independently derivable from pre-answer observations.

\textbf{Trajectory refinement and dataset finalization.}
For trajectories insufficiently grounded in tool observations, we introduced a retry stage that enforced additional tool use, followed by re-evaluating the resulting prediction. If a trajectory still failed to produce the correct answer, we used Qwen3.5-27B to repair the final rationale and prediction while preserving the original tool invocations and observations. Concretely, the model was provided with the initial trajectory, retrieved evidence, and ground-truth label, and was asked to generate a corrected rationale consistent with the evidence and target answer. We then merged all validated and repaired trajectories to construct the final distillation dataset, containing 54,249 training samples.

\section{Methodology}

\subsection{Genome-Conditioned Tool-Augmented Agent}

\textbf{Tool construction for genome-conditioned reasoning.}
We equip the agent with two biological tools: a similarity-based RAG tool and a GEM-based perturbation tool. The RAG tool is built from the synthetic SFT database containing field-level annotations for 8,000 strains. For each query strain, we retrieve the top-3 nearest strains in the gene-embedding space measured using cosine similarity and return their ground-truth annotations as tool observations. 
To prevent label leakage, the query strain's own ground-truth annotations are excluded during SFT data distillation.

For the GEM tool, we reconstruct strain-specific genome-scale metabolic models from protein sequences using CarveMe~\citep{machado2018fast}. Each strain is instantiated with three biomass templates, corresponding to Gram-positive bacteria, Gram-negative bacteria, and archaea, with the archaeal template adapted from iAF692~\citep{feist2006modeling, king2016bigg}. We evaluate each model under six medium perturbations: removing sulfate ($\mathrm{SO_4^{2-}}$), ferric iron ($\mathrm{Fe^{3+}}$), nitrate ($\mathrm{NO_3^-}$), nitrite ($\mathrm{NO_2^-}$), oxygen ($\mathrm{O_2}$), or all five components jointly. These perturbations probe major biochemical and ecological constraints, including sulfur assimilation, iron-dependent redox metabolism, nitrate/nitrite-associated nitrogen and respiratory pathways, and oxygen-dependent respiration. For each biomass template and perturbation, we compute the minimal medium supporting at least 10\% of the maximum growth rate, yielding 18 GEM-derived observations per strain for downstream physiological reasoning.

\textbf{Genome-token modal fusion.}
To incorporate genomic information into the language model, we adopt a lightweight genome-token fusion module to condition the language model on strain-level genomic information. Following multimodal LLM alignment pipelines~\citep{li2023blip, liu2023visual, wang2025biobridge}, we use LucaOne as a frozen genome encoder and employ a two-layer MLP projector $\phi(\cdot)$ to map the resulting genome representation $g_i$ into the LLM token-embedding space. The projected vector $z_i=\phi(g_i)$ is inserted as a continuous genome token alongside the textual instruction, tool observations, and task query, enabling genomic conditioning without architectural modifications to the backbone LLM.

\textbf{Agentic supervised fine-tuning.}
After genome-token alignment, we perform supervised fine-tuning to distill tool-augmented physiological reasoning into a compact agent. Using the constructed strain-centric corpus, we generate instruction trajectories with Qwen3.5-27B as the teacher model, where each trajectory contains the genome token, task query, tool observations, tool-use reasoning process, and final answer. We fine-tune Qwen3.5-4B as the student agent under a next-token prediction objective with a context length of 16,384 tokens. Throughout agentic SFT, LucaOne and the modal-fusion MLP are frozen, and only the student LLM is updated, preserving the genome-token alignment while adapting the model to tool-conditioned life-boundary prediction.

\subsection{Gene-Grounded Tool-Use Policy Optimization}

Because teacher trajectories are generated without genome embeddings, SFT primarily teaches tool syntax and evidence formatting. We therefore evaluate whether subsequent RL changes tool-selection behavior as a function of genome embeddings. We further optimize it with GRPO~\citep{shao2024deepseekmath}. Tool calls are treated as policy actions, while tool observations are masked from the policy loss.

\textbf{Sequence-level reward.}
For each sampled trajectory $\tau_i$, we assign a composite reward $r(\tau_i)=\lambda_{\mathrm{json}}r_{\mathrm{json}}+\lambda_{\mathrm{corr}}r_{\mathrm{corr}}+\lambda_{\mathrm{tool}}r_{\mathrm{tool}}+\lambda_{\mathrm{nt}}r_{\mathrm{nt}}$. 
The JSON reward $r_{\mathrm{json}}$ enforces that the final response is a single valid JSON object containing exactly the target field. 
The correctness reward $r_{\mathrm{corr}}$ evaluates biological accuracy using field-specific criteria: interval targets are rewarded for covering the ground-truth boundary while remaining compact, scalar targets are scored by normalized error, categorical and morphological targets are evaluated after label canonicalization, and substrate-source targets are scored by top-$k$ set overlap. 
The tool-use reward $r_{\mathrm{tool}}$ encourages evidence acquisition according to a scheduled tool-call target, while $r_{\mathrm{nt}}$ penalizes incorrect direct answers that bypass tool observations.
The tool-use advantage $A_{i,t}^{\mathrm{tool}}$ is obtained by applying
the same group-relative normalization to per-step tool-call shaping
rewards $r_{\mathrm{tool}}$, and is assigned only to model-generated
tool-call tokens (the textual content of each tool invocation, excluding
environment-inserted observations).

\textbf{Group-relative advantage estimation.}
For each prompt, we sample $G$ trajectories and compute a group-relative advantage $A_i^{(g)}=(r(\tau_i^{(g)})-\mathrm{mean}_j r(\tau_i^{(j)}))/(\mathrm{std}_j r(\tau_i^{(j)})+\epsilon)$. 
This normalization compares candidate trajectories under the same biological query and avoids the need for a learned value function. 
The resulting sequence advantage is assigned to final-answer tokens, while tool-call tokens receive separate tool-use shaping rewards.

\textbf{Counterfactual gene-grounding reward.}
A central challenge in genome-conditioned reasoning is that the policy may generate plausible answers from language priors or tool observations without effectively using the genome embedding. 
To encourage genuine genomic conditioning, we compute a counterfactual gene-grounding reward through two teacher-forced forward passes over the same sampled completion. 
In the first pass, the model is conditioned on the real genome embedding $g_i$; in the second pass, $g_i$ is replaced with an ablated zero-gene embedding $\mathbf{0}$. 
For each final-answer token $y_{i,t}$, we compute the log-probability gap $\Delta_{i,t}^{\mathrm{gene}}=\log \pi_{\theta}(y_{i,t}\mid c_{i,t},g_i)-\log \pi_{\theta}(y_{i,t}\mid c_{i,t},\mathbf{0})$, where $c_{i,t}$ denotes the textual context, prior generated tokens, tool observations, and task query before token $t$. 
% The gene-grounding term is then defined as $A_{i,t}^{\mathrm{gene}}=\max(r_{\mathrm{corr}}(\tau_i),0)\cdot \mathrm{clip}(\Delta_{i,t}^{\mathrm{gene}},-c,c)\cdot \mathbf{1}[t\in\mathrm{answer}]$. 
The gene-grounding term is then defined as
$A_{i,t}^{\mathrm{gene}}=\max(r_{\mathrm{corr}}(\tau_i),0)\cdot
\mathrm{clip}(\Delta_{i,t}^{\mathrm{gene}},-c,c)\cdot \mathbf{1}[t\in\mathrm{answer}]$,
where $c>0$ is a fixed clipping constant that bounds the per-token
log-probability gap and prevents a small number of high-magnitude tokens
from dominating the advantage.
Thus, the model is rewarded only when the real genome embedding increases the likelihood of biologically correct answer tokens relative to the zero-gene counterfactual, preventing reinforcement of gene-dependent but incorrect predictions.

\textbf{Optimization objective.}
The final token-level advantage is 
$A_{i,t}=A_i\mathbf{1}[t\in\mathrm{answer}]
+\lambda_{\mathrm{tool}}A_{i,t}^{\mathrm{tool}}
+\lambda_{\mathrm{gene}}A_{i,t}^{\mathrm{gene}}$,
where the tool-use term is applied to model-generated tool-call tokens and the gene-grounding term is applied only to final-answer tokens.
Let $m_{i,t}^{\mathrm{pol}}$ be a policy mask that excludes environment-inserted tool observations.
Following GRPO, we optimize
\begin{align*}
\resizebox{\linewidth}{!}{$
\mathcal{J}_{\mathrm{GRPO}}(\theta)
=
\mathbb{E}_{q,\{\tau_i\}_{i=1}^{G}}
\left[
\frac{1}{G}\sum_{i=1}^{G}
\frac{1}{|\tau_i|}
\sum_{t=1}^{|\tau_i|}
m_{i,t}^{\mathrm{pol}}
\left(
\min\!\left[
\rho_{i,t}A_{i,t},
\mathrm{clip}(\rho_{i,t},1-\epsilon,1+\epsilon)A_{i,t}
\right]
-\beta \mathrm{KL}_{i,t}
\right)
\right].
$}
\end{align*}
where $\rho_{i,t}=\pi_{\theta}(y_{i,t}\mid c_{i,t},g_i)/
\pi_{\theta_{\mathrm{old}}}(y_{i,t}\mid c_{i,t},g_i)$ and 
$\mathrm{KL}_{i,t}=\mathbb{D}_{\mathrm{KL}}[
\pi_{\theta}(\cdot\mid c_{i,t},g_i)\Vert
\pi_{\mathrm{ref}}(\cdot\mid c_{i,t},g_i)]$.
This objective optimizes structured output validity, biological correctness, efficient tool use, and explicit dependence on the genome representation.
During RL, the LucaOne encoder and genome-fusion MLP remain frozen, and only the Qwen3.5-4B policy is updated.

\section{Experiments}
\subsection{Experimental Setup}
In this subsection, we detailed our experimental settings, with additional details in Appendix~\ref{app:experimental_setup}.

\textbf{Models.}
We build our gene-grounded agent using Qwen3.5-4B~\citep{yang2025qwen3,qwen35}, which serves as the language backbone for genome-modality fusion. We inject genome embeddings through the proposed fusion module and train the resulting agent with the three-stage pipeline described above. To contextualize the performance of our trained agent against stronger general-purpose LLMs, we additionally evaluate several larger frontier baselines, including DeepSeek-V3.2~\citep{liu2025deepseek32}, DeepSeek-R1~\citep{guo2025deepseekr1}, GLM-4.7~\citep{zeng2025glm}, and Kimi-K2~\citep{team2025kimi}. These models are used as strong text-based baselines for assessing whether explicit gene-grounded training provides benefits beyond scaling general language-model capacity.

\textbf{Implementation details.}
We train the model in three stages. We first conduct full-parameter gene-text instruction tuning from a gene-aligned Qwen3.5-4B checkpoint for one epoch, using sequence length 1,024, effective batch size 256, and learning rate $2\times10^{-5}$. We then perform agent SFT on distilled tool-use trajectories for one epoch, with sequence length 16,384, effective batch size 128, and learning rate $5\times10^{-6}$. Finally, we optimize the SFT agent with GRPO on BacDive-derived prompts, using a group size of 4 with a global rollout batch size of 64 and learning rate $1\times10^{-6}$. During RL, we allow at most 5 tool-calling iterations, freeze the genome-fusion module, and update only the language-policy parameters.

\textbf{Evaluation metrics.}
We use task-specific metrics for heterogeneous microbial traits. In the following experiments, we primarily report interval coverage rate (ICR) for physiological boundaries, RMSE for optimal conditions, mAP@5 for substrate-source traits, and accuracy for categorical physiological and morphological traits. 
% We also compute mean interval width and interval score for interval traits, micro-F1@5 for substrate traits, and macro-F1 after label canonicalization for categorical and morphological traits.

\subsection{Main Results}

\begin{table}[t]
\centering
\caption{
Main results on genome-conditioned microbial physiology prediction. We compare the base language model, genome--text modal fusion model, agent SFT model, and RL-optimized agent across physiological boundary, optimal-condition, metabolism, categorical physiology, and morphology tasks. Best results are shown in \textbf{bold}; second-best results are \underline{underlined}.
}
\label{tab:main_results}
\footnotesize
\setlength{\tabcolsep}{4pt}
\renewcommand{\arraystretch}{0.98}
\begin{tabular}{@{}llccccc@{}}
\toprule
\textbf{\shortstack[l]{Tasks}} 
& \textbf{Field} 
& \textbf{Metric}
& \textbf{Base Model} 
& \textbf{Fusion Model} 
& \textbf{SFT Agent} 
& \textbf{RL Agent} \\
\midrule

\multirow{3}{*}{\textbf{\shortstack[l]{Physiological\\Boundary}}}
& Growth temperature range & ICR $\uparrow$
& 0.157{\scriptsize$\pm$0.007} 
& 0.163{\scriptsize$\pm$0.094} 
& \underline{0.179}{\scriptsize$\pm$0.001} 
& \textbf{0.182}{\scriptsize$\pm$0.003} \\
& pH range & ICR $\uparrow$
& 0.287{\scriptsize$\pm$0.011} 
& 0.296{\scriptsize$\pm$0.146} 
& \underline{0.458}{\scriptsize$\pm$0.001} 
& \textbf{0.497}{\scriptsize$\pm$0.067} \\
& Salinity range & ICR $\uparrow$
& 0.466{\scriptsize$\pm$0.024} 
& 0.471{\scriptsize$\pm$0.089} 
& \underline{0.781}{\scriptsize$\pm$0.051} 
& \textbf{0.818}{\scriptsize$\pm$0.045} \\

\midrule
\multirow{3}{*}{\textbf{\shortstack[l]{Physiological\\Optimum}}}
& Optimal growth temperature & RMSE $\downarrow$
& 8.570{\scriptsize$\pm$0.233} 
& 8.449{\scriptsize$\pm$0.580} 
& \underline{6.449}{\scriptsize$\pm$0.093} 
& \textbf{6.373}{\scriptsize$\pm$0.113} \\
& Optimal pH & RMSE $\downarrow$
& 1.124{\scriptsize$\pm$0.040} 
& 1.077{\scriptsize$\pm$0.383} 
& \underline{0.736}{\scriptsize$\pm$0.000} 
& \textbf{0.726}{\scriptsize$\pm$0.017} \\
& Optimal salinity & RMSE $\downarrow$
& 3.526{\scriptsize$\pm$0.074} 
& 3.441{\scriptsize$\pm$0.080} 
& \underline{2.553}{\scriptsize$\pm$0.014} 
& \textbf{2.532}{\scriptsize$\pm$0.029} \\

\midrule
\multirow{4}{*}{\textbf{Metabolism}}
& Carbon source & mAP@5 $\uparrow$
& 0.000{\scriptsize$\pm$0.067} 
& 0.000{\scriptsize$\pm$0.000} 
& \underline{0.040}{\scriptsize$\pm$0.007} 
& \textbf{0.054}{\scriptsize$\pm$0.022} \\
& Electron acceptor & mAP@5 $\uparrow$
& 0.003{\scriptsize$\pm$0.048} 
& 0.003{\scriptsize$\pm$0.006} 
& \underline{0.233}{\scriptsize$\pm$0.018} 
& \textbf{0.249}{\scriptsize$\pm$0.018} \\
& Electron donor & mAP@5 $\uparrow$
& 0.000{\scriptsize$\pm$0.086} 
& 0.000{\scriptsize$\pm$0.000} 
& \underline{0.095}{\scriptsize$\pm$0.014} 
& \textbf{0.118}{\scriptsize$\pm$0.027} \\
& Nitrogen source & mAP@5 $\uparrow$
& 0.003{\scriptsize$\pm$0.032} 
& 0.003{\scriptsize$\pm$0.002} 
& \underline{0.209}{\scriptsize$\pm$0.001} 
& \textbf{0.217}{\scriptsize$\pm$0.012} \\

\midrule
\multirow{4}{*}{\textbf{\shortstack[l]{Categorical\\Physiology}}}
& Carbon fixation pathway & Accuracy $\uparrow$
& 0.081{\scriptsize$\pm$0.042} 
& 0.097{\scriptsize$\pm$0.037} 
& \underline{0.414}{\scriptsize$\pm$0.009} 
& \textbf{0.419}{\scriptsize$\pm$0.000} \\
& Energy type & Accuracy $\uparrow$
& 0.047{\scriptsize$\pm$0.155} 
& 0.042{\scriptsize$\pm$0.031} 
& \underline{0.500}{\scriptsize$\pm$0.000} 
& \textbf{0.505}{\scriptsize$\pm$0.008} \\
& Oxygen tolerance & Accuracy $\uparrow$
& 0.271{\scriptsize$\pm$0.029} 
& 0.244{\scriptsize$\pm$0.145} 
& \underline{0.541}{\scriptsize$\pm$0.008} 
& \textbf{0.549}{\scriptsize$\pm$0.012} \\
& Photosynthesis type & Accuracy $\uparrow$
& 0.168{\scriptsize$\pm$0.118} 
& 0.263{\scriptsize$\pm$0.094} 
& \underline{0.611}{\scriptsize$\pm$0.011} 
& \textbf{0.625}{\scriptsize$\pm$0.016} \\

\midrule
\multirow{5}{*}{\textbf{Morphology}}
& Cell shape & Accuracy $\uparrow$
& 0.684{\scriptsize$\pm$0.024} 
& 0.670{\scriptsize$\pm$0.353} 
& \underline{0.800}{\scriptsize$\pm$0.002} 
& \textbf{0.802}{\scriptsize$\pm$0.003} \\
& Flagella & Accuracy $\uparrow$
& 0.133{\scriptsize$\pm$0.121} 
& 0.125{\scriptsize$\pm$0.075} 
& \underline{0.536}{\scriptsize$\pm$0.006} 
& \textbf{0.542}{\scriptsize$\pm$0.004} \\
& Gram stain & Accuracy $\uparrow$
& 0.529{\scriptsize$\pm$0.133} 
& 0.681{\scriptsize$\pm$0.380} 
& \underline{0.853}{\scriptsize$\pm$0.000} 
& \textbf{0.855}{\scriptsize$\pm$0.003} \\
& Motility & Accuracy $\uparrow$
& 0.597{\scriptsize$\pm$0.033} 
& 0.659{\scriptsize$\pm$0.301} 
& \underline{0.700}{\scriptsize$\pm$0.002} 
& \textbf{0.701}{\scriptsize$\pm$0.000} \\
& Spore formation & Accuracy $\uparrow$
& 0.699{\scriptsize$\pm$0.057} 
& 0.803{\scriptsize$\pm$0.337} 
& \underline{0.838}{\scriptsize$\pm$0.000} 
& \textbf{0.839}{\scriptsize$\pm$0.002} \\
\bottomrule
\end{tabular}
\vspace{-0.4cm}
\end{table}

\textbf{GGBound shows a consistent progression from passive genome conditioning to selective, genome-grounded reasoning.}
Table~\ref{tab:main_results} compares the base model, genome--text fusion model, SFT agent, and RL-optimized GGBound across all benchmark tasks. The RL agent achieves the best performance on every target, while the SFT agent is consistently second-best, indicating that agentic training is the main driver of performance gains. These results suggest that life-boundary prediction requires more than access to genomic representations; it also benefits from learning how to acquire, filter, and synthesize biological evidence during generation. 

\textbf{Genome--text fusion provides limited standalone gains, whereas agentic SFT produces the largest transition in capability.}
Modal fusion improves several continuous targets over the base model, including optimal growth temperature and optimal salinity, suggesting that frozen LucaOne embeddings encode useful environmental-preference signals. However, fusion remains unreliable on many categorical and morphology tasks, where performance often lags behind the base model. In contrast, agentic SFT substantially improves nearly all task groups, raising salinity-range ICR from $0.471$ to $0.781$, electron acceptor mAP@5 from $0.003$ to $0.233$, oxygen-tolerance accuracy from $0.244$ to $0.541$. This indicates that distilled trajectories help the model convert genomic and tool-derived evidence into standardized physiological predictions.

\textbf{Reinforcement learning yields smaller but systematic gains over a strong SFT agent.}
Compared with SFT, the RL agent improves all targets, including pH-range ICR from $0.458$ to $0.497$, salinity-range ICR from $0.781$ to $0.818$, optimal pH RMSE from $0.736$ to $0.726$, and nitrogen-source mAP@5 from $0.209$ to $0.217$. Gains are also consistent across categorical and morphology tasks, such as photosynthesis type ($0.611$ to $0.625$), carbon fixation pathway ($0.414$ to $0.419$), and flagella ($0.536$ to $0.542$). Although these improvements are often modest, their consistency across heterogeneous targets supports the role of GRPO and the counterfactual gene-grounding reward in refining biologically grounded predictions rather than merely changing output format.

\begin{wraptable}{r}{0.295\textwidth}
\centering
\small
\vspace{-0.6cm}
\caption{Averaged tool-use.}
\setlength{\tabcolsep}{2pt}  % 缩小列间距，提高紧凑性
\label{tab:avg_tool}
\begin{tabular}{lccc}
\toprule
\textbf{Model} & \textbf{RAG} & \textbf{GEM} & \textbf{Avg.} \\
\midrule
Base & 1.052 & 5.914 & 6.966 \\
Fusion & 1.024 & 5.632 & 6.656 \\
SFT Agent & 0.994 & 0.580 & 1.574 \\
RL Agent & 0.993 & 0.739 & 1.732 \\
\bottomrule
\vspace{-0.5cm}
\end{tabular}
\end{wraptable}
\textbf{Tool-use statistics indicate that better performance comes from more selective evidence acquisition rather than more tool calls.}
As shown in Table~\ref{tab:avg_tool}, the base and fusion models invoke substantially more tools on average, largely due to frequent GEM calls, but still underperform the agentic models. Agentic SFT sharply reduces average tool usage while improving performance, suggesting that trajectory supervision teaches a more efficient evidence-acquisition policy. RL slightly increases tool usage relative to SFT, especially GEM calls, while achieving the best overall accuracy, indicating that policy optimization improves when and how biological tools are used. Overall, the results suggest that genome-token fusion provides an initial grounding signal, agentic SFT establishes effective tool-conditioned reasoning, and counterfactual gene-grounded RL further calibrates the policy toward accurate strain-level physiological prediction.

\subsection{Strain-Name Priors and Frontier LLM Comparison}

\begin{figure}[t]
    \centering
    \includegraphics[width=\linewidth]{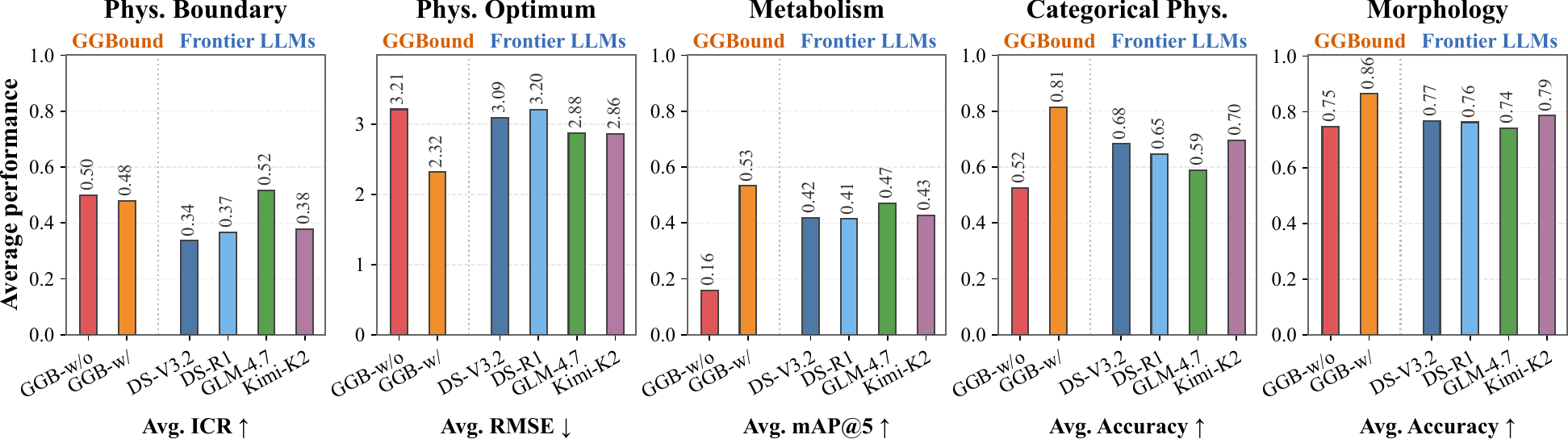}
    \caption{Average performance under strain-name evaluation. GGB-w/o and GGB-w/ denote GGBound without and with strain-name input, respectively. Frontier LLMs are directly prompted with strain names, while GGBound uses genome conditioning and GEM access.}
    \label{fig:avg_compare}
    \vspace{-0.35cm}
\end{figure}

\textbf{Strain-name input provides a complementary source of taxonomic prior knowledge.}
Figure~\ref{fig:avg_compare} compares GGBound with and without strain-name information against frontier LLMs queried directly with strain names. This setting differs from the anonymized benchmark because strain names expose species- or taxon-level cues, allowing models to exploit biological knowledge stored during pretraining. For GGBound, we retain genome-conditioned inference and allow GEM tool calls; for frontier LLMs, we directly query each model using the strain name and target field, since these models may already encode substantial microbial knowledge internally. (Appendix~\ref{app:frontier_llm_strain_name} and~\ref{app:strain_name_tool}).

\textbf{Name-based priors substantially improve GGBound on most physiological categories.}
Adding strain names improves GGBound on physiological optima, metabolism, categorical physiology, and morphology. The largest gains appear on metabolism and categorical physiology, where average mAP@5 increases from $0.16$ to $0.53$ and average accuracy increases from $0.52$ to $0.81$, respectively. These gains suggest that strain names activate useful taxonomic priors that complement genome-derived and GEM-derived evidence. The only exception is physiological-boundary prediction, where average ICR slightly decreases from $0.50$ to $0.48$, indicating that name-based priors can sometimes conflict with genome-conditioned or tool-derived signals for heterogeneous interval annotations.

\textbf{GGBound remains competitive with frontier LLMs even when those models receive strain names.}
Despite the favorable setting for frontier LLMs, GGBound with strain-name input achieves the best average performance on physiological optima, metabolism, categorical physiology, and morphology. It outperforms the strongest frontier LLM average on metabolism ($0.53$ vs. $0.47$), categorical physiology ($0.81$ vs. $0.70$), and morphology ($0.86$ vs. $0.79$), while achieving the lowest average RMSE on physiological optima ($2.32$). GLM-4.7 performs best on physiological-boundary ICR, suggesting that boundary ranges may be more strongly influenced by broad taxonomic priors or interval-calibration effects. Overall, this comparison shows that strain names can substantially raise performance by activating internal biological knowledge, while GGBound further benefits from integrating such priors with genome-conditioned reasoning and GEM-based evidence.

% \subsection{Ablation Study}

\section{Conclusion, Broader Impacts, and Limitations}
\label{sec:limitations_impacts}

\textbf{Conclusion.}
We introduced \textsc{GGBound}, a genome-grounded, tool-augmented LLM agent for microbial life-boundary prediction. We formulate strain-level physiology prediction as a unified genome-to-physiology task and construct a strain-centric benchmark covering physiological boundaries, optima, metabolic capabilities, categorical traits, and morphology. \textsc{GGBound} combines LucaOne genome embeddings with RAG and GEM tool observations, and is trained through gene--text alignment, agentic SFT, and GRPO with a counterfactual gene-grounding reward. Results show that agentic training substantially improves genome-conditioned prediction, while reinforcement learning further refines biologically grounded outputs.

\textbf{Broader Impacts.}
Genome-grounded life-boundary prediction can help prioritize cultivation conditions, reduce wet-lab screening cost, and support microbial ecology and biotechnology. The proposed benchmark also provides a standardized setting for evaluating whether biological agents reason from genomic evidence rather than taxonomic memorization. Model outputs should nevertheless be treated as hypotheses, not experimental instructions, and downstream use should follow expert review and appropriate biosafety standards.

\textbf{Limitations.}
A key limitation is the highly compressed genome representation used for modality fusion. Each strain is represented by a single LucaOne-derived embedding projected into the LLM token space, which may lose substantial gene-level, pathway-level, and structural information needed for fine-grained physiological prediction. This bottleneck is particularly problematic for substrate utilization, metabolic capability, and stress-response traits, where localized genomic evidence may be essential. Future work should explore richer genome-to-LLM interfaces, including multiple genome tokens, gene-set-aware pooling, and adaptive retrieval over gene features.

\newpage
\bibliographystyle{unsrtnat}
\bibliography{reference}

%%%%%%%%%%%%%%%%%%%%%%%%%%%%%%%%%%%%%%%%%%%%%%%%%%%%%%%%%%%%

\newpage
\appendix

\section{Dataset and Benchmark Construction}
\label{app:dataset_benchmark}

\subsection{IJSEM Extraction Pipeline}
\label{app:ijsem_extraction}

We constructed the IJSEM-derived strain-physiology corpus through a four-stage LLM-assisted extraction pipeline. The four stages correspond to: (i) evidence retrieval from full IJSEM articles, (ii) identification of primary and comparison strains, (iii) bounded strain-level trait extraction, and (iv) LLM-based extraction quality scoring. The goal of this pipeline is not to ask an LLM to directly hallucinate structured biological labels from article titles or taxonomic names, but to progressively convert verbatim article evidence into auditable strain-level records. Each stage is constrained by explicit-only extraction rules and produces intermediate artifacts that can be inspected independently.

\textbf{Input articles.}
The raw input consists of IJSEM articles represented as structured JSON files containing title, abstract, main body, figures, tables, and paper links when available. For each article, we concatenate the abstract and main text. If tables are available, we append table content to the article input in a marked table block. This is important because IJSEM taxonomic descriptions often report strain-level physiology, growth ranges, substrate utilization, and morphology in tables rather than narrative paragraphs.

\textbf{Stage 1: evidence retrieval from full articles.}
The first stage performs coarse evidence retrieval. Given the full article text and extracted tables, the LLM is instructed to retrieve only text that reports intrinsic strain- or species-level factual information. The output of this stage is not yet a structured database record. Instead, the model returns verbatim article spans, including paragraphs and tables, that contain information relevant to strain characterization.

The retrieval target includes species names, strain identifiers, type-strain markers, isolation source, habitat, geographic location, culture collection identifiers, genome or sequence accessions, growth conditions, physiological traits, biochemical properties, substrate utilization, carbon source profiles, electron donors and acceptors, enzyme activities, and metabolic capabilities. Growth medium or medium composition is retained when it is biologically informative, for example when it indicates which substrates, nutrients, or energy sources the strain can use.

We use strict retention rules to prevent loss of attribution context. If a paragraph contains any strain- or species-level physiological, biochemical, metabolic, or growth property, the entire paragraph is retained verbatim. The model is explicitly prohibited from extracting only individual sentences from such paragraphs. Similarly, if a table contains any information about the target strain or species, the entire table is retained, including caption, body, row and column labels, and footnotes. This design avoids fragmenting table-derived evidence and preserves the mapping between strain labels and properties. If no relevant biological content is found, the model returns a fixed null marker, and the article is excluded from downstream extraction.

\textbf{Stage 2: strain identification and inventory construction.}
The second stage operates only on the Stage-1 retrieved evidence. Its purpose is to identify all biological strains mentioned in the article and separate them into two groups: \texttt{paper\_strains}, which are the strains proposed, described, or directly studied by the paper, and \texttt{comparison\_strains}, which are reference or comparison strains used for taxonomic, phylogenetic, phenotypic, or biochemical comparison.

The Stage-2 prompt requires the model to output the most complete strain names explicitly supported by the article. The model may combine genus, species, nomenclatural qualifier, and strain identifier only when all components are explicitly stated in the text or table captions. It is not allowed to infer missing taxonomy, expand abbreviations without explicit support, or treat placeholders as strains unless the article explicitly defines them as strain names. Table captions are given priority when they define full strain names or mappings between abbreviated table labels and strains.

We also apply a conservative deduplication rule. If multiple mentions refer to the same strain identifier, only the most complete explicitly supported name is retained. For example, if both a bare strain identifier and a full genus--species--strain expression appear, the full expression is retained. Nomenclatural qualifiers such as \textit{sp. nov.}, \textit{gen. nov.}, and \textit{comb. nov.} are preserved at this stage because they are part of the source article's taxonomic statement. The output of this stage is a strict JSON object with unique \texttt{paper\_strains} and \texttt{comparison\_strains} lists.

\textbf{Stage 3: bounded strain-level trait extraction.}
The third stage extracts structured strain-level properties. Unlike Stage 2, which identifies strains, Stage 3 is bounded by the \texttt{paper\_strains} list. The model is given the Stage-1 evidence and the Stage-2 primary strain list, and is instructed to extract properties only for those primary strains. It is not allowed to create new strains, rename strains, or extract benchmark labels for comparison strains. This bounded design is critical for preventing properties of reference organisms from being incorrectly assigned to the newly described strains.

For every primary strain, the model outputs a JSON entry keyed by the exact strain name or identifier from Stage 2. If no properties are explicitly available for a strain, the output still contains the strain identifier and an empty auxiliary field. This ensures that the structured output remains aligned with the strain inventory.

The extraction schema covers environmental origin, genome and sequence information, morphology, growth boundaries, optimal conditions, metabolism, and other strain-intrinsic traits. The target fields include habitat, latitude and longitude, water depth, environment type, sampling site, sampling depth, sampling temperature, sampling pressure, geographic region, taxonomy, top-hit taxon, sequence similarity, genome accession, 16S rRNA accession or sequence, GC content, genome size, plasmid presence, reported functional genes, cell shape, cell size, motility, flagella, spore formation, Gram stain, photosynthesis type, photopigments, light conditions, medium type, isolation medium, isolation temperature, enrichment pressure, enrichment time, growth temperature optimum, growth temperature range, pH optimum, pH range, salinity optimum, salinity range, growth pressure, doubling time, energy type, electron donor, electron acceptor, carbon fixation pathway, carbon source, nitrogen source, oxygen tolerance, antibiotic sensitivity, antibiotic resistance, major fatty acids, and respiratory quinones.

The extractor follows strict attribution rules. A property is assigned to a strain only if the Stage-1 evidence explicitly links that property to the strain. The linkage may appear locally in a sentence, through a table column, through a caption-defined strain mapping, or through a clear contextual scope. Global or group-level statements are not assigned to individual strains unless the article explicitly supports such assignment. Values are copied verbatim, including symbols, units, ranges, punctuation, and qualifiers, except for minimal character-level repair of obvious encoding artifacts introduced by PDF or HTML conversion.

The prompt also includes field-disambiguation rules to avoid biologically invalid mappings. For example, budding pattern, reproduction mode, hyphae formation, or filamentous stage must not be mapped to motility or flagella. Temperatures associated with storage or preservation must not be mapped to isolation temperature. Fungal LSU or ITS accessions must not be forced into bacterial 16S rRNA fields. Explicitly stated properties that do not fit the predefined schema are stored in an \texttt{extra} dictionary rather than forced into an unrelated field.

\textbf{Stage 4: LLM-based quality scoring.}
The fourth stage evaluates the Stage-3 structured extraction against the Stage-1 retrieved evidence. The rater is given the original retrieved evidence and the structured JSON output. It is instructed to ignore any natural-language summary field and score only the structured strain-level entries under \texttt{strains}. The rater evaluates extraction quality along six dimensions: strain coverage completeness, field coverage completeness, attribution correctness, value fidelity to the source text, schema compliance, and hallucination/noise control.

The rater also produces evidence-backed diagnostics. Missing expected fields must be supported by short verbatim quotes from the evidence. Incorrect or unsupported extracted values must either include a corrected source-supported value or be marked as unsupported. This stage is designed to detect incomplete extraction, incorrect strain-property attribution, unsupported values, schema violations, and hallucinated fields.

\textbf{Quality filtering.}
Only records whose Stage-4 quality score is greater than 7 are retained for subsequent corpus and benchmark construction. This threshold removes records with weak coverage, ambiguous attribution, schema inconsistency, or potential hallucination. As a result, the benchmark is constructed from records that satisfy three constraints: the source evidence was retrieved verbatim in Stage 1, the strain identity was explicitly identified in Stage 2, and the structured trait extraction passed the Stage-4 quality filter.

\textbf{Expert review of coarse retrieval.}
After completing the Stage-1 coarse IJSEM retrieval, domain experts manually reviewed the retrieved evidence blocks. The review focused on whether the extracted paragraphs and tables were faithful to the original articles, whether they contained relevant strain- or species-level biological information, and whether the model introduced hallucinated or paraphrased content. The experts reported that the reviewed Stage-1 outputs were accurate. This expert review supports the use of Stage-1 retrieved text as the evidence base for the downstream strain-identification and trait-extraction stages.

\subsection{Prompt Templates for IJSEM Extraction}
\label{app:ijsem_prompts}

To improve reproducibility, we provide the core prompt templates used in the IJSEM extraction pipeline. The prompts are stage-specific and enforce conservative extraction behavior. All stages require strict JSON or verbatim-text outputs and prohibit unsupported biological inference.

\textbf{Stage-1 evidence retrieval prompt.}
The Stage-1 prompt retrieves relevant article evidence before any structured extraction. The core system instruction is:

\begin{quote}\small
You are an information extraction agent for microbiology strain characterization papers. Your task is to extract sentences, paragraphs, and table-related text, including full tables, that report key intrinsic, strain- or species-level factual information. This includes characteristics of the organism itself, not procedural protocols.

If a paragraph reports any strain- or species-level physiological, biochemical, metabolic, or growth property, it must be extracted verbatim in full, even if it also contains experimental methods, assay names, media descriptions, or protocol-like language. If a table contains any information about the target strain or species, extract the entire table, including all rows, columns, captions, and footnotes.

Extract only what is explicitly stated in the article. Do not infer, assume, generalize, or add information. Output only the extracted text, verbatim. Remove HTML tags. Deduplicate identical content. Do not add explanations, labels, summaries, rewriting, paraphrasing, or reordering. If no relevant content is found, output exactly \texttt{NO\_RELEVANT}.
\end{quote}

The corresponding user prompt asks the model to extract key strain- and species-level factual information from the article, focusing on species names, strain identifiers, isolation source, geographic location, growth or metabolic properties, culture collection identifiers, NCBI-related accessions, and full tables that report strain or species properties.

\textbf{Stage-2 strain identification prompt.}
The Stage-2 prompt converts retrieved evidence into a strain inventory. The core instruction is:

\begin{quote}\small
You are a strain extraction system for microbiology articles. Given verbatim text from a paper, including paragraphs, tables, and captions, extract all biological strains mentioned in the paper. Output the most complete strain names by combining explicitly stated genus, species, and strain identifiers from the text or captions. Classify strains into strains proposed or studied by the paper and strains used for comparison.

You may combine taxonomy and strain only if all parts are explicitly stated. Do not infer or guess missing information. Table captions have priority for full strain names. Copy names verbatim, including symbols, punctuation, and type-strain markers. Ignore row numbers, labels, and placeholders. Nomenclatural qualifiers such as \textit{sp. nov.}, \textit{gen. nov.}, or \textit{comb. nov.} are part of the explicit species name when they appear in the text and must be retained.

If multiple names refer to the same strain identifier, retain only the most complete name. The most complete name is the one that includes the maximum explicitly stated taxonomic information. Output strict JSON only:
\[
\{\texttt{"paper\_strains"}: [...], \texttt{"comparison\_strains"}: [...]\}.
\]
\end{quote}

This prompt allows the pipeline to distinguish primary strains from comparison organisms before any trait label is extracted.

\textbf{Stage-3 bounded trait extraction prompt.}
The Stage-3 prompt performs structured extraction only for \texttt{paper\_strains}. The core instruction is:

\begin{quote}\small
You are a strict strain-level information extraction agent for microbiology papers. Given original evidence text and a list of paper strains, extract explicitly stated strain-level properties for each listed strain and generate a concise summary. Extract only information explicitly stated in the evidence. Do not infer, guess, normalize, calculate, or add missing facts. Copy values verbatim, except for minimal repair of clearly corrupted symbols caused by text encoding. Do not create new strains or rename strains. Ignore any strain not listed in \texttt{PAPER\_STRAINS}.

Assign a property to a strain only if the text explicitly links it to that strain, for example through a local statement, a table column, or a caption-defined mapping. Global or group-level statements must not be assigned to individual strains. Unstated fields must be omitted. If an explicitly stated property does not semantically match any target field, store it in \texttt{extra} rather than forcing it into an unrelated field.

Output strict JSON only.
% \[
% \{
% \texttt{"summary"}: \ldots,
% \texttt{"strains"}: \{
% \texttt{"<strain>"}: \{
% \texttt{"strain\_name\_or\_id"}: \texttt{"<strain>"},
% \ldots,
% \texttt{"extra"}: \{\}
% \}
% \}
% \}.
% \]
\end{quote}

The prompt lists all allowed fields and field-specific disambiguation rules. This prevents common errors such as mapping budding pattern to motility, preservation temperature to isolation temperature, or fungal barcode accessions to bacterial 16S rRNA accession fields.

\textbf{Stage-4 quality scoring prompt.}
The Stage-4 prompt audits the structured extraction. The core instruction is:

\begin{quote}\small
You are a strict evaluator for strain-level information extraction. Given original evidence text and extracted JSON, evaluate accuracy, completeness, attribution, and schema compliance. The top-level summary field may be present but must be ignored for scoring, coverage checks, attribution, and diagnostics. Only content under \texttt{strains} is subject to evaluation.

A value is correct only if it is explicitly stated or directly supported by combining information explicitly stated in the evidence. Inference beyond the text, guessing, normalization, or unit conversion is not allowed. Strain-level facts must be explicitly tied to that strain either locally in text or through clear contextual linkage such as table caption, table column, or section scope. Global or group-level statements must not be assigned to strains.

Score all dimensions: strain coverage completeness, field coverage completeness, attribution correctness, value fidelity, schema compliance, and hallucination/noise control. For every reported issue, provide evidence using a short verbatim quote or \texttt{NO\_EXPLICIT\_SUPPORT\_FOUND}. Output strict JSON only.
\end{quote}

We use the resulting quality score to filter records and retain only entries with score greater than 7 for benchmark construction.

\subsection{Strain Name Normalization and Cross-Database Matching}
\label{app:strain_matching}

The IJSEM extraction stages preserve strain names in their article-level form. This is necessary for traceability because taxonomic articles often include nomenclatural qualifiers, type-strain markers, strain aliases, collection identifiers, and punctuation that should not be discarded during evidence extraction. However, direct lookup in NCBI and BacDive requires a standardized query representation. We therefore apply a separate strain-name normalization stage for cross-database matching. The normalized fields are used only for database lookup and alignment; they do not replace the original IJSEM evidence string.

\textbf{Normalization target.}
For each raw strain name, the normalizer outputs a strict JSON object with two fields:
\[
\texttt{NCBI\_scientific\_name}
\quad\text{and}\quad
\texttt{NCBI\_modifier}.
\]
The scientific-name field is normalized to the binomial form ``Genus species''. The normalizer keeps only the first valid binomial name and removes nomenclatural qualifiers such as \textit{sp. nov.} and \textit{gen. nov.}, subspecies annotations, strain annotations, parenthetical content, and collection-code strings. The strain modifier field is extracted separately as the first strain-like token appearing after the species name.

\textbf{Modifier extraction.}
The modifier is intended to capture the article's primary strain identifier rather than a culture collection alias. A valid modifier may contain letters, numbers, hyphens, and type-strain notation. Trailing type-strain suffixes such as \texttt{T} and \texttt{(T)} are removed. Text after equality signs is ignored to avoid mixing the primary strain identifier with equivalent culture collection deposits. Collection identifiers such as DSM, KCTC, MCCC, NCTC, ATCC, JCM, CGMCC, CICC, and NBRC are ignored as primary modifiers. If no valid strain-like modifier is explicitly present, the modifier field is left empty. The normalizer is explicitly prohibited from guessing or inferring missing identifiers.

\textbf{Example.}
For an article-level strain name such as
\[
\text{``\textit{Guyparkeria halopsychrophila} sp. nov. LHSS19-1(T)''},
\]
the normalized scientific name is
\[
\texttt{Guyparkeria halopsychrophila},
\]
and the normalized modifier is
\[
\texttt{LHSS19-1}.
\]
The original article string remains stored in the IJSEM-derived record, while the normalized fields are used for NCBI search.

\textbf{NCBI matching.}
We first use the normalized scientific name and modifier to retrieve NCBI genome and protein records. Matches are retained only when the species-level identity and strain-level modifier support the same strain identity. This prevents a species-level match from being incorrectly treated as a strain-level match. Ambiguous cases, missing-modifier cases, collection-code-only matches without sufficient strain support, and records whose taxonomic identity cannot be reconciled are excluded from exact-match benchmark construction.

\textbf{BacDive matching.}
After NCBI matching, we further align strains to BacDive phenotype annotations using standardized scientific names and strain identifiers. BacDive provides structured phenotype records that complement IJSEM-derived literature evidence. The matching stage produces a unified strain-centric record containing: (i) original IJSEM evidence, (ii) normalized strain identity, (iii) NCBI genome and protein information, (iv) BacDive phenotype annotations when available, and (v) downstream genome representations. Conservative matching is used throughout to reduce false positives, particularly when multiple strains share a species name but differ in strain identifiers or culture collection deposits.

\textbf{Separation between evidence preservation and database lookup.}
We distinguish between evidence strings and normalized lookup keys. The original IJSEM strain name is preserved for traceability and citation to the article. The normalized NCBI fields are used only to query external databases. This separation allows the pipeline to preserve the taxonomic wording of IJSEM descriptions while still enabling reliable cross-database alignment.

\subsection{Trait Normalization and Controlled Vocabularies}
\label{app:trait_normalization}

The Stage-3 extractor preserves values as they appear in the source article. These raw values are useful for traceability but are not directly suitable for benchmark evaluation because IJSEM articles vary in wording, units, abbreviations, and reporting granularity. We therefore apply trait normalization after extraction and quality filtering. The goal is to map explicitly supported raw values into controlled benchmark targets while retaining the original evidence strings.

\textbf{General normalization principle.}
Normalization is performed only when the raw value is explicitly supported by the source evidence and can be mapped unambiguously to the benchmark schema. We do not infer missing labels from taxonomy, habitat, related species, or medium recipes. Values that are biologically informative but incompatible with the benchmark schema are retained in the corpus as auxiliary information but are not used as benchmark labels.

\textbf{Physiological boundary traits.}
For growth temperature range, pH range, and salinity range, explicitly reported lower and upper bounds are converted into standardized interval targets. Temperature is represented in degrees Celsius. pH is represented on the original pH scale. Salinity is standardized when the source unit permits reliable conversion. If the article reports only qualitative descriptions such as ``halophilic'', ``acidophilic'', or ``thermotolerant'' without numeric bounds, the statement is retained in the raw corpus but not used as an interval benchmark label.

\textbf{Optimal-condition traits.}
For optimal growth temperature, optimal pH, and optimal salinity, explicitly reported optima are converted into scalar targets. If an article reports an optimum as a narrow range, the range is retained in the corpus and converted to a scalar only when needed by the evaluation metric. If multiple optima are reported under different media or conditions, the value is used only when the target condition is clearly attributable to the strain-level optimum; otherwise, it is treated as ambiguous and excluded from the benchmark target set.

\textbf{Substrate-source and metabolic traits.}
For carbon source, electron donor, electron acceptor, and nitrogen source, raw extracted values are mapped to controlled label sets. We canonicalize obvious aliases, spelling variants, and chemically equivalent surface forms while preserving the original source strings. Multi-label traits are represented as sets. Medium composition is handled conservatively: a component of a growth or isolation medium is not automatically treated as a positive substrate-utilization label unless the article explicitly states that the strain utilizes, requires, oxidizes, reduces, or grows with that component in the relevant metabolic role. This avoids converting every ingredient in a complex medium into a positive metabolic capability.

\textbf{Categorical physiological traits.}
For oxygen tolerance, energy type, photosynthesis type, and carbon fixation pathway, extracted descriptions are mapped to controlled vocabularies with domain-expert guidance. Oxygen tolerance labels are assigned only when the article explicitly describes aerobic, anaerobic, facultative, microaerophilic, or aerotolerant behavior. Energy type and photosynthesis labels are assigned only when the evidence supports the corresponding physiological category. Carbon fixation pathway labels are assigned only when the pathway or its diagnostic genes or reactions are explicitly stated. Ambiguous descriptions are retained in raw fields but excluded from closed-form benchmark labels unless they can be mapped unambiguously.

\textbf{Morphological traits.}
For morphology, we standardize cell shape, motility, flagella, spore formation, and Gram stain. These fields are normalized separately because taxonomic descriptions often mention them in adjacent sentences but they are not logically interchangeable. The presence of flagella is not automatically mapped to motility unless motility is explicitly reported. Conversely, non-motility is not automatically mapped to absence of flagella. Budding pattern, filamentous growth, hyphae formation, reproductive mode, and related morphological observations are not mapped to motility or flagellar state. They are retained as auxiliary properties when explicitly stated.

\textbf{Canonicalization and exclusion.}
For each candidate benchmark label, we require three conditions: (i) the value is explicitly supported by the source evidence, (ii) the Stage-4 extraction quality score for the record is greater than 7, and (iii) the value can be mapped to the target schema without unsupported inference. This conservative policy favors precision over recall. It reduces noisy labels in the benchmark, while retaining richer raw information for future analyses.

\subsection{Benchmark Construction and Filtering}
\label{app:benchmark_filtering}

After IJSEM extraction, cross-database matching, trait normalization, and genome alignment, we construct a strain-centric benchmark for genome-conditioned microbial life-boundary prediction. Each benchmark instance pairs a strain-level genome representation with a query for one target physiological attribute. A single strain may contribute multiple benchmark instances when it has multiple high-quality, explicitly supported labels.

\textbf{Record-level filtering.}
We first filter IJSEM-derived records using the Stage-4 extraction quality score. Only records with a quality score greater than 7 on the 0--10 scale are retained. This threshold is applied before trait-specific instance construction, ensuring that all benchmark instances derived from IJSEM evidence originate from records that pass the extraction-quality filter. This step removes low-confidence structured extractions with weak attribution, incomplete field coverage, schema violations, or potentially unsupported values.

\textbf{Trait-level filtering.}
Within retained records, each candidate trait label is further checked for benchmark compatibility. A value is included as a benchmark label only if it satisfies three conditions: (i) it is explicitly supported by the source evidence, (ii) it maps unambiguously to one benchmark field, and (iii) it matches the expected output type of that field. Numeric growth ranges are used for physiological-boundary prediction; numeric optima are used for optimal-condition prediction; controlled substrate/source labels are used for metabolic prediction; and controlled categorical or morphological labels are used for classification tasks. Raw values that are explicit but not compatible with the benchmark target format are preserved in the full corpus but excluded from benchmark evaluation.

\textbf{Genome alignment requirement.}
Benchmark strains must be alignable to genomic information. We therefore require a successful cross-database match to NCBI genome or protein records sufficient for constructing the strain-level genome representation. Strains with high-quality physiological labels but no usable genomic match are retained in the broader corpus but excluded from genome-conditioned benchmark evaluation.

\textbf{Instance construction.}
For each eligible strain record, we create one benchmark instance for every non-null target field. Each instance contains an anonymized strain identifier, a genome representation, a target field name, a task-specific query, and the normalized ground-truth label. The benchmark is therefore organized as a collection of strain--field pairs rather than as one record per strain. This design supports heterogeneous prediction targets while preserving a unified genome-conditioned input format.

\textbf{Benchmark groups.}
We organize benchmark instances into five task groups: physiological boundary prediction, optimal-condition prediction, substrate/source prediction, categorical physiological trait prediction, and morphology prediction. Physiological boundary tasks include growth temperature range, pH range, and salinity range. Optimal-condition tasks include optimal growth temperature, optimal pH, and optimal salinity. Substrate/source tasks include carbon source, electron donor, electron acceptor, and nitrogen source. Categorical physiological tasks include oxygen tolerance, energy type, photosynthesis type, and carbon fixation pathway. Morphology tasks include cell shape, motility, flagella, spore formation, and Gram stain.

\textbf{Target spaces.}
Table~\ref{tab:benchmark_target_spaces} summarizes the benchmark subtasks and their output spaces. Boundary and optimum tasks use numeric lower--upper objects. Categorical physiology and morphology tasks use controlled single-label vocabularies, except for Boolean-style traits, which are represented as true/false outputs. Substrate/source tasks are represented as multi-label lists over controlled vocabularies.

\begin{table}[h]
\centering
\caption{Benchmark subtasks and output spaces. Boundary and optimum tasks use numeric lower--upper objects; categorical and morphology tasks use controlled single-label or Boolean outputs; substrate/source tasks use controlled multi-label outputs.}
\scriptsize
\begin{tabular}{p{2.5cm}p{3.15cm}p{6.5cm}}
\toprule
Category & Subtask & Output space \\
\midrule

Physiological boundary
& \texttt{growth\_temperature\_range\_C}
& Numeric interval in Celsius: \texttt{\{"lower": number, "upper": number\}} \\

Physiological boundary
& \texttt{pH\_range}
& Numeric interval: \texttt{\{"lower": number, "upper": number\}} \\

Physiological boundary
& \texttt{salinity\_range}
& Numeric interval in \% w/v: \texttt{\{"lower": number, "upper": number\}} \\

\midrule

Optimal condition
& \texttt{growth\_temperature\_opt\_C}
& Numeric optimum in Celsius, represented as \texttt{\{"lower": number, "upper": number\}} \\

Optimal condition
& \texttt{pH\_opt}
& Numeric optimum, represented as \texttt{\{"lower": number, "upper": number\}} \\

Optimal condition
& \texttt{salinity\_opt\_wv\_percent}
& Numeric optimum in \% w/v, represented as \texttt{\{"lower": number, "upper": number\}} \\

\midrule

Categorical physiology
& \texttt{oxygen\_tolerance}
& \texttt{aerobic}, \texttt{anaerobic}, \texttt{facultative\_anaerobic}, \texttt{microaerophilic}, \texttt{aerotolerant\_anaerobic} \\

Categorical physiology
& \texttt{energy\_type}
& \texttt{chemoorgano}, \texttt{chemolitho}, \texttt{photoautotrophic}, \texttt{photoheterotrophic} \\

Categorical physiology
& \texttt{photosynthesis\_type}
& \texttt{oxygenic}, \texttt{anoxygenic}, \texttt{none}, \texttt{strictly aerobic} \\

Categorical physiology
& \texttt{carbon\_fixation\_pathway}
& \texttt{Calvin-Benson-Bassham}, \texttt{reverse\_tca}, \texttt{wood-ljungdahl}, \texttt{3\_hydroxypropionate\_4\_hydroxybutyrate}, \texttt{rump}, \texttt{serine\_cycle}, \texttt{none}, \texttt{heterotrophic} \\

\midrule

Morphology
& \texttt{cell\_shape}
& \texttt{coccoid}, \texttt{rod}, \texttt{short\_rod}, \texttt{spiral}, \texttt{pleomorphic} \\

Morphology
& \texttt{motility}
& Boolean output: \texttt{true} or \texttt{false} \\

Morphology
& \texttt{flagella}
& \texttt{absent}, \texttt{polar}, \texttt{peritrichous}, \texttt{lophotrichous} \\

Morphology
& \texttt{spore\_formation}
& Boolean output: \texttt{true} or \texttt{false} \\

Morphology
& \texttt{gram\_stain}
& \texttt{positive}, \texttt{negative}, \texttt{variable} \\

\midrule

Substrate/source
& \texttt{carbon\_source}
& Multi-label list from: \texttt{inorganic\_carbon}, \texttt{C1}, \texttt{sugars}, \texttt{alcohols\_polyols}, \texttt{organic\_acids\_TCA}, \texttt{amino\_acids}, \texttt{complex\_media}, \texttt{hydrocarbons}, \texttt{aromatics}, \texttt{none\_or\_negative}, \texttt{unknown}, \texttt{other}, \texttt{hydrogen} \\

Substrate/source
& \texttt{electron\_donor}
& Multi-label list from: \texttt{hydrogen}, \texttt{reduced\_sulfur}, \texttt{iron\_or\_metal}, \texttt{organic}, \texttt{hydrocarbons}, \texttt{aromatics}, \texttt{unknown}, \texttt{sugars}, \texttt{organic\_acids\_TCA}, \texttt{alcohols\_polyols}, \texttt{C1}, \texttt{alcohols}, \texttt{complex\_media}, \texttt{inorganic\_carbon}, \texttt{amino\_acids} \\

Substrate/source
& \texttt{electron\_acceptor}
& Multi-label list from: \texttt{oxygen}, \texttt{nitrate\_nitrite}, \texttt{sulfur\_oxyanions\_or\_S0}, \texttt{fumarate\_or\_other\_organic}, \texttt{iron\_manganese}, \texttt{arsenate\_antimonate\_selenate}, \texttt{carbon\_dioxide}, \texttt{none}, \texttt{unknown}, \texttt{organic} \\

Substrate/source
& \texttt{nitrogen\_source}
& Multi-label list from: \texttt{ammonium}, \texttt{nitrate\_nitrite}, \texttt{urea}, \texttt{amino\_acids\_peptides}, \texttt{yeast\_extract}, \texttt{nitrogen\_fixation}, \texttt{other}, \texttt{unknown}, \texttt{none\_or\_negative}, \texttt{complex\_media} \\

\bottomrule
\end{tabular}
\label{tab:benchmark_target_spaces}
\end{table}

\textbf{Evaluation-time anonymization.}
To evaluate genome-grounded prediction rather than memorization from taxonomic names, strain names are anonymized during benchmark evaluation. The model receives the genome-derived representation and the target query, but not the true species or strain name. This prevents models from exploiting memorized associations between known microbial names and common phenotypes. We separately evaluate a strain-name setting as an auxiliary analysis, while the primary benchmark uses anonymized strain identities.

\textbf{Leakage control.}
We apply leakage control at both the strain and tool levels. Benchmark samples are kept disjoint from downstream agent-training data. During tool-augmented inference and trajectory construction, the query strain's own ground-truth annotation is not exposed through retrieval outputs. For the similarity-based RAG tool, the target strain itself is excluded from retrieval. This prevents direct label leakage while still allowing the agent to use information from biologically related strains as external evidence. The separation between raw IJSEM evidence, normalized database identifiers, training records, and benchmark instances is maintained throughout construction.

\subsection{Human Expert Validation}
\label{app:expert_validation}

In addition to automated LLM-based quality scoring, we conducted human expert validation after the Stage-1 IJSEM evidence retrieval. This validation was performed before the downstream strain-identification and trait-extraction stages. The purpose was to verify that the coarse retrieval step was faithful to the original articles, because all later stages operate on the retrieved evidence rather than the full article.

\textbf{Validation criteria.}
Experts assessed three aspects of the Stage-1 outputs. First, they checked factual fidelity: whether the extracted paragraphs and tables were copied from the source article without hallucinated content. Second, they checked relevance: whether the retrieved text contained intrinsic strain- or species-level information relevant to microbial characterization. Third, they checked context preservation: whether physiology-rich paragraphs and relevant tables were retained completely enough to support downstream strain-property attribution.

\textbf{Outcome.}
The experts reported that the reviewed Stage-1 outputs were accurate and faithful to the original IJSEM articles. In particular, the retrieved blocks preserved relevant biological information and did not introduce unsupported facts. Based on this review, we used the Stage-1 outputs as the evidence base for Stage 2 and Stage 3. The subsequent Stage-4 automatic rater provides an additional structured quality-control layer at the record level, and only records with quality score greater than 7 are retained for benchmark construction.

\text{Role of expert validation.}
The expert review serves a different role from the Stage-4 rater. Expert validation checks whether the coarse evidence retrieval step produces faithful source evidence. Stage-4 scoring checks whether the structured JSON extraction correctly attributes fields to strains and complies with the target schema. Combining these two checks reduces both upstream evidence-retrieval errors and downstream structured-extraction errors.

\subsection{Summary of Data Construction Decisions}
\label{app:data_construction_summary}

The complete data construction pipeline follows a conservative evidence-to-label design. First, we retrieve relevant evidence from IJSEM articles verbatim. Second, we identify primary paper strains and comparison strains from the retrieved evidence. Third, we extract structured strain-level properties only for the primary paper strains. Fourth, we score the structured records and retain only records with quality score greater than 7. Fifth, we normalize strain names for NCBI and BacDive matching while preserving the original IJSEM evidence strings. Finally, we harmonize explicit trait values into controlled benchmark labels and construct genome-conditioned task instances only when a strain has both a high-quality label and a usable genomic match.

This design intentionally favors precision over recall. Some explicit but ambiguous values are retained in the broader corpus but excluded from the benchmark. Similarly, strains with physiological evidence but insufficient genome matching are not used for genome-conditioned evaluation. These filtering decisions improve the reliability of the benchmark and reduce the risk of evaluating models against noisy or weakly attributed labels.

\section{Experimental Setup}
\label{app:experimental_setup}

\subsection{Tool construction for genome-conditioned reasoning}
To support genome-conditioned life-boundary prediction, we equip the agent with two biological tools: a similarity-based RAG tool and a GEM metabolic perturbation tool. 
The RAG tool is built from the synthetic SFT database, which contains field-level annotations for 8,000 strains. 
Given a query strain, we compute its similarity to database strains in the gene-embedding space and return the top-3 nearest strains with their ground-truth field annotations as tool observations. 
During SFT data distillation, we exclude the target strain's own ground-truth annotations from the tool outputs to prevent label leakage.

For the GEM tool, we reconstruct strain-specific genome-scale metabolic models from protein sequences using CarveMe~\citep{machado2018fast}. 
To account for major taxonomic and cell-envelope differences in biomass requirements, we instantiate each strain with three biomass templates: Gram-positive bacteria, Gram-negative bacteria, and archaea, with the archaeal template adapted from iAF692\footnote{\url{http://bigg.ucsd.edu/models/iAF692}}~\citep{feist2006modeling, king2016bigg}. 
We then simulate each model under six medium perturbation conditions. 
Specifically, five perturbations individually remove sulfate ($\mathrm{SO_4^{2-}}$), ferric iron ($\mathrm{Fe^{3+}}$), nitrate ($\mathrm{NO_3^-}$), nitrite ($\mathrm{NO_2^-}$), or oxygen ($\mathrm{O_2}$) from the medium, while the sixth removes all five components simultaneously. 
For each biomass template and perturbation condition, we compute the minimal medium composition supporting at least 10\% of the maximum growth rate, yielding 18 GEM-derived tool observations per strain. 
These observations provide mechanistic evidence about nutrient dependence, redox feasibility, and metabolic constraints for downstream physiological reasoning.

\subsection{Tool Interfaces}
\label{app:tool_interfaces}

GGBound is trained and evaluated as a genome-conditioned tool-using agent. Each prompt contains a special \texttt{<gene>} token, which is replaced internally by the projected genome representation, and an anonymized genome handle that can be passed to external tools. The model does not observe the true strain name, raw genome identifier, or any physical storage path during anonymized evaluation.

The environment exposes two tools: a retrieval tool and a GEM tool. Both tools take the anonymized genome handle as input and return structured textual observations that are appended to the conversation. Tool observations are visible to the model in subsequent turns, but they are treated as environment outputs rather than model-generated tokens during training.

\textbf{Retrieval tool.}
The retrieval tool provides nearest-neighbor phenotypic evidence. Its input is the anonymized genome handle of the query strain. Its output is a compact JSON-style observation containing the top retrieved strains in genome-embedding space, their similarity scores, and a small set of phenotype fields relevant to the current prediction task. A typical observation has the following form:
\begin{tcolorbox}[
  colback=gray!3,
  colframe=gray!35,
  boxrule=0.3pt,
  arc=1pt,
  left=4pt,
  right=4pt,
  top=4pt,
  bottom=4pt
]
{\scriptsize\ttfamily
\{\par
\quad "tool": "rag\_tool",\par
\quad "top\_similar\_records": [\par
\quad\quad \{\par
\quad\quad\quad "rank": 1,\par
\quad\quad\quad "similarity": 0.87,\par
\quad\quad\quad "phenotypes": \{\par
\quad\quad\quad\quad "growth\_temperature\_range\_C": \{"lower": 20, "upper": 37\},\par
\quad\quad\quad\quad "pH\_range": \{"lower": 6.0, "upper": 8.0\},\par
\quad\quad\quad\quad "oxygen\_tolerance": "aerobe",\par
\quad\quad\quad\quad "gram\_stain": "negative",\par
\quad\quad\quad\quad "carbon\_source": ["glucose", "acetate"]\par
\quad\quad\quad \}\par
\quad\quad \}\par
\quad ],\par
\quad "retrieved\_count": 3\par
\}
}
\end{tcolorbox}
The retrieval output is intended to provide contextual evidence from biologically similar strains. Since retrieved neighbors may be incomplete or locally misleading, the model is instructed to treat the observation as evidence rather than as a direct label oracle.

\textbf{GEM tool.}
The \texttt{gem\_tool} provides physiological and metabolic-model evidence from genome-scale metabolic models (GEMs) under predefined biomass formulations and medium-perturbation settings. Its input consists of the anonymized genome handle and a configuration id. The 18 configurations correspond to three biomass formulations---archaeal, Gram-negative, and Gram-positive---crossed with six perturbation conditions. The perturbations remove one of \(\mathrm{O_2}\), \(\mathrm{Fe^{3+}}\), \(\mathrm{NO_3^-}\), \(\mathrm{NO_2^-}\), or \(\mathrm{SO_4^{2-}}\), or remove all five simultaneously.

The tool output is a structured observation containing a minimal substrate dictionary predicted by the corresponding GEM configuration. The dictionary maps exchange-reaction identifiers to the uptake amounts required by the model under the selected physiological-metabolic setting. These raw GEM outputs are not treated as final phenotype labels; instead, they provide mechanistic evidence that the agent can use when reasoning about carbon sources, electron acceptors, nitrogen sources, and related metabolic traits.

A typical observation is:
\begin{tcolorbox}[
  colback=gray!3,
  colframe=gray!35,
  boxrule=0.3pt,
  arc=1pt,
  left=4pt,
  right=4pt,
  top=4pt,
  bottom=4pt
]
{\scriptsize\ttfamily
\{\par
\quad "tool": "gem\_tool",\par
\quad "configuration\_id": 7,\par
\quad "minimal\_substrate\_dict": \{\par
\quad\quad "EX\_3amp\_e": 1.25,\par
\quad\quad "EX\_arg\_\_L\_e": 3.18,\par
\quad\quad "EX\_so4\_e": 0.03,\par
\quad\quad "EX\_xylan8\_e": 28.16,\par
\quad\quad "EX\_zn2\_e": 0.002\par
\quad \},\par
\quad "error": null\par
\}
}
\end{tcolorbox}
The full tool response may contain additional exchange reactions and real-valued uptake coefficients. If the requested GEM configuration is unavailable, the tool returns the same schema with an empty \texttt{minimal\_substrate\_dict} and a non-null error message. This fixed output format allows the agent to reason over successful and failed GEM calls uniformly.

The final assistant answer is still required to follow the task-specific canonical output schema. We do not evaluate predictions against raw exchange-reaction identifiers; GEM observations serve only as intermediate physiological-metabolic evidence.

\textbf{Agent-tool interaction.}
At each rollout step, the model may either call one of the tools or produce a final answer. When the model emits a tool call, the environment executes the call and appends the returned observation to the conversation. The model then continues generation conditioned on the original prompt, previous assistant turns, and accumulated tool observations. This loop terminates when the model emits a final answer or reaches the maximum number of tool-calling rounds.

% The final answer must be exactly one JSON object containing the requested target field. For example, for a carbon-source query, the final answer has the form:
% \begin{verbatim}
% {"carbon_source": ["glucose", "acetate"]}
% \end{verbatim}
% Intermediate turns may contain tool calls or concise reasoning, but the final answer cannot contain extra prose, Markdown fences, or fields unrelated to the requested target.

\subsection{Generating Modality-Fused Instruction Data}
We construct a modality-fused instruction dataset from structured biological strain QA pairs. For each selected strain, the pipeline preserves the original attribute-level QA pairs and augments them with two generated data types: paraphrased attribute QA pairs and holistic strain-level description QA pairs. For attribute-level augmentation, each valid QA pair is rewritten by a language model while preserving its original meaning. The prompt explicitly requires factual consistency, no new information, and semantic equivalence between the original and rewritten answers. To improve linguistic diversity, the prompt varies clause order, syntactic structure, and information flow. Near-duplicate rewrites are filtered using normalized sequence similarity against both the original QA pair and previously accepted variants. For holistic augmentation, all QA pairs of a strain are first converted into a fact block. The model then generates a short global question, such as asking for an overall description of the strain, followed by a comprehensive answer grounded only in the provided facts. This converts isolated attributes into integrated descriptions covering identity, taxonomy, habitat, sampling source, location, and other available strain information.
Generation is performed with batched vLLM inference. The system enforces JSON-formatted outputs, retries malformed generations, and supports incremental JSONL writing with resume functionality. The final dataset contains unified records with \texttt{id}, \texttt{name}, \texttt{question}, and \texttt{answer}, combining fine-grained attribute supervision with global strain-level descriptive supervision.

\subsection{Agentic SFT Data Collection Details}
\label{app:agentic_sft_data}

We construct the agentic SFT corpus using a multi-stage trajectory distillation pipeline. The goal is to distill tool-using biological reasoning trajectories from a strong teacher model into supervised fine-tuning examples for the compact genome-conditioned agent. Each training example contains an anonymized strain embedding handle, a biological trait prediction query, optional intermediate tool calls, tool observations, and a final JSON-formatted answer.

\textbf{Teacher rollout generation.}
We use Qwen3.5-27B as the teacher model to generate initial multi-turn trajectories with vLLM. For each input sample, the teacher is prompted as a biological reasoning agent with access to \texttt{rag\_tool} and \texttt{gem\_tool}. The input pool is constructed from BacDive-derived SFT records. Each prompt specifies the target trait field, an anonymized embedding handle, available tool schemas, and the required final JSON answer format.

For each sample, we generate four candidate trajectories using temperature \(1.0\), top-\(p=0.95\), maximum generation length 4096, maximum model context length 16,384, and up to six tool-calling rounds. The teacher may call \texttt{rag\_tool} to retrieve phenotypically annotated nearest-neighbor strains and \texttt{gem\_tool} to obtain genome-scale metabolic model evidence under selected biomass and medium-perturbation configurations. The resulting trajectories contain a sequence of assistant turns, tool calls, tool observations, intermediate planning text, and a final structured answer.

\textbf{Teacher system prompt.}
The teacher system prompt defines the model as an expert biology agent for microbial trait prediction. It introduces two biological tools:
\begin{itemize}
    \item \texttt{rag\_tool}, which retrieves similar strains from the genome-embedding database and returns compact phenotype annotations;
    \item \texttt{gem\_tool}, which returns genome-scale metabolic model evidence for a selected biomass formulation and perturbation condition.
\end{itemize}
The prompt also defines the \texttt{gem\_tool} identifier space. Tool ids 1--6 correspond to archaeal biomass formulations, 7--12 to Gram-negative biomass formulations, and 13--18 to Gram-positive biomass formulations. Within each block of six, the offset specifies the unavailable medium component: \(\mathrm{O_2}\), \(\mathrm{Fe^{3+}}\), \(\mathrm{NO_3^-}\), \(\mathrm{NO_2^-}\), \(\mathrm{SO_4^{2-}}\), or all five components jointly. The teacher is instructed to prefer conclusions grounded in tool evidence, avoid unsupported guesses, gather complementary evidence when useful, avoid unnecessary duplicate calls, and produce exactly one final JSON object.

\textbf{User prompt format.}
For each subtask, the user prompt contains the prediction problem, the anonymized embedding handle, tool-use guidance, and the target-specific final answer schema. The prompt follows the template:

\begin{tcolorbox}[
  colback=gray!3,
  colframe=gray!35,
  boxrule=0.3pt,
  arc=1pt,
  left=4pt,
  right=4pt,
  top=4pt,
  bottom=4pt
]
{\scriptsize\ttfamily
Problem: <task-specific trait prediction instruction>\par
\medskip
Anonymous genome handle: <anonymous genome handle>\par
\medskip
Tool-use guidance:\par
You have access to rag\_tool and gem\_tool using the anonymous genome handle above. Use these tools when they provide useful evidence, and prefer tool-grounded answers over unsupported prior beliefs.\par
\medskip
Guidelines:\par
1. rag\_tool provides retrieval-grounded phenotype evidence, but may be incomplete or noisy.\par
2. gem\_tool provides complementary GEM evidence and may help verify or refine retrieval evidence.\par
3. When both tools provide relevant evidence, synthesize them.\par
4. If one tool returns weak, empty, redundant, or irrelevant evidence, rely more on the stronger evidence.\par
5. Multiple tool calls are allowed when they improve confidence, but avoid unnecessary repeats.\par
6. If tool evidence conflicts, choose the most defensible answer.\par
7. Use the model's tool-calling format for tool calls and keep intermediate turns concise.\par
\medskip
When ready, output exactly one final JSON object matching: <schema>.\par
Only the final answer must follow this JSON schema; intermediate reasoning or tool-calling turns need not be JSON.
}
\end{tcolorbox}
For interval and optimum-condition tasks, the prompt additionally encourages the teacher to use GEM evidence when it can strengthen or validate retrieval-based predictions, rather than relying only on nearest-neighbor annotations.

\textbf{Subtask coverage.}
The final distillation corpus contains 54,249 examples across 17 realized subtasks. The distribution is shown in Table~\ref{tab:sft_distill_task_counts}.

\begin{table}[h]
\centering
\small
\caption{Subtask distribution of the final agentic SFT distillation corpus.}
\begin{tabular}{llr}
\toprule
Category & Subtask & Number of examples \\
\midrule
Growth interval & \texttt{growth\_temperature\_range\_C} & 7948 \\
Growth interval & \texttt{pH\_range} & 3481 \\
Growth interval & \texttt{salinity\_range} & 3894 \\
Growth optimum & \texttt{growth\_temperature\_opt\_C} & 4074 \\
Growth optimum & \texttt{pH\_opt} & 3251 \\
Growth optimum & \texttt{salinity\_opt\_wv\_percent} & 2317 \\
Physiology & \texttt{oxygen\_tolerance} & 5805 \\
Physiology & \texttt{energy\_type} & 305 \\
Morphology & \texttt{cell\_shape} & 5218 \\
Morphology & \texttt{motility} & 5139 \\
Morphology & \texttt{flagella} & 648 \\
Morphology & \texttt{spore\_formation} & 2794 \\
Morphology & \texttt{gram\_stain} & 5448 \\
Substrate/source & \texttt{carbon\_source} & 3572 \\
Substrate/source & \texttt{electron\_donor} & 46 \\
Substrate/source & \texttt{electron\_acceptor} & 78 \\
Substrate/source & \texttt{nitrogen\_source} & 231 \\
\midrule
Total & -- & 54249 \\
\bottomrule
\end{tabular}
\label{tab:sft_distill_task_counts}
\end{table}

\textbf{Trajectory parsing and scoring.}
After teacher rollout generation, we parse the final answer of each sampled trajectory and compute both strict and best-effort scores. The strict parser requires the final assistant response to be exactly one valid JSON object containing the requested target field. The best-effort parser is used only offline to recover candidate predictions from otherwise noisy generations. Each parsed trajectory is scored using the same family of biological correctness functions as in RL: interval coverage and compactness for boundary traits, normalized error for optimum traits, canonical-label match for categorical and morphology traits, and top-\(k\) set overlap for substrate-source traits.

For each input sample, we select the best trajectory among the teacher candidates using a field-aware ranking procedure. The primary criterion is biological correctness. Ties are broken using strict JSON validity, parse success, field-aware tool evidence quality, fewer tool errors, richer non-error tool interaction, and shorter final answers. For interval and optimum prediction tasks, trajectories that use both \texttt{rag\_tool} and \texttt{gem\_tool} are preferred over retrieval-only trajectories when answer quality is comparable, because these tasks benefit from complementary metabolic feasibility evidence.

\textbf{Retry and evidence strengthening.}
The distillation pipeline includes two retry stages to improve weak, incomplete, or retrieval-only trajectories. The first retry targets imperfect trajectories that used \texttt{rag\_tool} but did not use \texttt{gem\_tool}. The second retry applies a field-aware policy: interval and optimum tasks are retried whenever the selected trajectory remains imperfect; substrate-source tasks are retried when the trajectory is imperfect and lacks GEM evidence; and other tasks are retried mainly for retrieval-only misses. The forced-retry prompt requires the teacher to call \texttt{rag\_tool} first and then use at least one or two \texttt{gem\_tool} calls before finalization, depending on the target field. During the second merge, 20,843 retry candidates were compared against the original selected trajectories, and 5,207 original trajectories were replaced by stronger retry trajectories.

\textbf{Cleaning and finalization.}
For selected trajectories whose final answer still mismatches the ground-truth label, we construct long-context cleaning prompts. The cleaner receives the full prior tool trace, the original final answer, the parsed prediction, the ground-truth label, and a task-specific difference summary. It is instructed to minimally edit only the final assistant answer while preserving the existing tool calls and tool observations. The cleaner is not allowed to invent new tool outputs or alter the evidence trace.

We use this cleaning stage to repair the supervised final-answer target, not to create synthetic tool observations. When a mismatch is repaired, the corrected final JSON answer is embedded back into the original trajectory. The preceding tool trace is preserved so that the SFT data still reflects the actual evidence available before the final answer. This separates tool-use behavior imitation from final-answer supervision and avoids training the model on fabricated observations.

\textbf{Final SFT format.}
The final agentic SFT dataset contains 54,249 records. Each record stores the true genome-embedding path in a top-level metadata field used by the training collator, while the conversation shown to the model contains only the anonymized embedding handle. Each record includes:
\begin{itemize}
    \item \texttt{id}: anonymized strain identifier;
    \item \texttt{name}: species-level name used as metadata;
    \item \texttt{gene\_cls\_path}: true \texttt{cls\_emb.npy} path used by the collator, not exposed in the conversation;
    \item \texttt{system}: biology-agent system prompt;
    \item \texttt{tools}: JSON schemas for \texttt{rag\_tool} and \texttt{gem\_tool};
    \item \texttt{conversations}: ShareGPT-style multi-turn trajectory containing the user prompt, assistant tool calls, tool observations, intermediate assistant turns, and final JSON answer.
\end{itemize}
The final construction consists of 29,914 cleaned embedded trajectories and 24,335 original trajectories that were already correct after selection and retry. During SFT, only model-generated assistant content is optimized; environment-provided tool observations are retained as context but are not treated as assistant-generated targets.

\subsection{Reward and Optimization Details}
\label{app:reward_optimization}

We optimize the genome-aware tool-use policy with Group Relative Policy Optimization (GRPO). For each prompt, the sampler generates multiple completions, executes any requested tools, scores each completed trajectory with task-specific rewards, normalizes rewards within the prompt group, and applies a clipped policy-gradient update to model-generated tokens. Tool observations inserted by the environment are masked from the policy loss.

\paragraph{Training configuration.}
The GRPO stage starts from the agentic supervised-finetuned checkpoint. Training prompts are constructed from BacDive-derived records and contain the system instruction, the \texttt{<gene>} placeholder, the target trait field, the anonymized genome handle, tool-use guidance, and the required final JSON schema. Table~\ref{tab:grpo_training_config} summarizes the main counterfactual GRPO configuration. Since the genome-projection module is frozen, the RL stage updates only the language-model policy.

\begin{table}[ht]
\centering
\small
\caption{Main hyperparameters for the counterfactual GRPO stage.}
\begin{tabular}{lll}
\toprule
Category & Hyperparameter & Value \\
\midrule
Sampling & Number of generations per prompt & 4 \\
Sampling & Temperature & 1.0 \\
Sampling & Top-\(p\) & 0.95 \\
Sampling & Top-\(k\) & 50 \\
\midrule
Optimization & Epochs & 1 \\
Optimization & Per-device batch size & 8 \\
Optimization & Gradient accumulation steps & 1 \\
Optimization & Steps per generation & 1 \\
Optimization & Learning rate & \(1\times 10^{-6}\) \\
Optimization & Learning-rate schedule & Cosine \\
Optimization & Warmup ratio & 0.03 \\
Optimization & GRPO clipping parameter & 0.2 \\
Optimization & Reference KL coefficient & 0.0 \\
\midrule
Context and tools & Maximum completion length & 8192 \\
Context and tools & Maximum vLLM generation tokens & 2048 \\
Context and tools & Maximum tool-calling rounds & 5 \\
\bottomrule
\end{tabular}
\label{tab:grpo_training_config}
\end{table}

\textbf{Rollout procedure.}
For each training row, the sampler repeats the same prompt \(\texttt{num\_generations}\) times. In the main configuration, each prompt therefore has four sampled completions. During generation, vLLM receives both the tokenized prompt and the genome modality:
\begin{verbatim}
{
  "prompt_token_ids": ids,
  "multi_modal_data": {"gene": gene_array}
}
\end{verbatim}
If a sampled trajectory contains tool calls, the environment executes the calls, appends the resulting tool messages, and resumes generation from the expanded conversation. This process continues until a final answer is produced or the maximum number of tool-calling iterations is reached. The model-generated assistant tokens are optimized, while environment-provided tool payloads are treated as observations and excluded from the policy-loss mask.

\textbf{Sequence-level rewards.}
The trajectory-level reward combines strict output formatting, biological correctness, tool-use shaping, and a penalty for incorrect direct answers without tool use:
\[
R_{\mathrm{seq}}
=
w_{\mathrm{json}}R_{\mathrm{json}}
+
w_{\mathrm{corr}}R_{\mathrm{corr}}
+
w_{\mathrm{tool}}R_{\mathrm{tool}}
+
w_{\mathrm{nt}}R_{\mathrm{no\_tool}}
+
R_{\mathrm{external}}.
\]
In the main counterfactual GRPO configuration, the default weights are:
\[
w_{\mathrm{json}}=0.5,\quad
w_{\mathrm{corr}}=1.0,\quad
w_{\mathrm{tool}}=1.0,\quad
w_{\mathrm{nt}}=1.0.
\]

The JSON-format reward is strict:
\[
R_{\mathrm{json}} =
\begin{cases}
+1, & \text{if the final answer is exactly one valid JSON object with the target field},\\
-1, & \text{otherwise}.
\end{cases}
\]
Malformed JSON, Markdown fences, extra prose, missing target fields, wrong target fields, null values, or additional unsupported fields receive \(-1\).

The biological correctness reward \(R_{\mathrm{corr}}\) depends on the target trait type. Interval-valued traits, such as growth temperature range, pH range, and salinity range, are rewarded for covering the ground-truth interval while remaining compact. Optimal-condition traits are scored by normalized absolute error. Categorical physiological and morphology traits receive \(+1\) for an exact match after label canonicalization and \(-1\) otherwise. Substrate-source traits, including carbon source, electron donor, electron acceptor, and nitrogen source, are evaluated using top-\(k\) micro-F1 with \(k=5\):
\[
R_{\mathrm{corr}} = 2\cdot \mathrm{MicroF1@5} - 1.
\]
Thus a perfect substrate prediction receives \(+1\), while a prediction with no overlap receives \(-1\).

\textbf{Tool-use reward.}
The tool-use reward encourages the policy to acquire evidence while discouraging both zero-tool shortcuts and excessive tool calls. Let \(c\) be the number of tool calls in a trajectory and \(t\) be the target number of calls. The target is annealed over training:
\[
t = t_{\mathrm{init}} + p(t_{\mathrm{final}} - t_{\mathrm{init}}),
\]
where \(p\in[0,1]\) denotes training progress. In the main configuration,
\[
t_{\mathrm{init}}=4.0,\qquad t_{\mathrm{final}}=2.0.
\]
The tool-use reward is:
\[
R_{\mathrm{tool}} =
\begin{cases}
-1.0, & c=0,\\
0.25 + 0.75\sqrt{c/t}, & 0 < c \leq t,\\
\max(0.25, 1.0 - 0.5(c-t)/t), & c > t.
\end{cases}
\]
This schedule makes nonzero tool use substantially better than no tool use, but does not reward unbounded tool calling.

\textbf{Counterfactual gene-usage reward.}
The counterfactual trainer adds a token-level reward that encourages the final answer to depend on the genome embedding. For each sampled completion, the trainer computes teacher-forced token log-probabilities twice: once with the real genome embedding \(g\), and once with an ablated zero genome embedding. For answer token \(i\), we define:
\[
\Delta_i =
\log \pi_\theta(y_i \mid x, g)
-
\log \pi_\theta(y_i \mid x, 0).
\]
The gap is clipped to
\[
[-c_{\mathrm{gene}}, c_{\mathrm{gene}}],
\]
where the default cap is \(c_{\mathrm{gene}}=5.0\). With positive-only gating enabled, the token-level gene-usage reward is:
\[
R_{\mathrm{gene},i}
=
\max(R_{\mathrm{corr}},0)
\cdot
\mathrm{clip}(\Delta_i, -c_{\mathrm{gene}}, c_{\mathrm{gene}})
\cdot
\mathbb{1}[i\in \mathrm{answer}].
\]
The main configuration uses \texttt{gene\_usage\_reward\_weight}=0.5. This reward is applied only to final-answer tokens and not to tool-call payloads or environment-inserted tool observations. The correctness gate prevents the model from being rewarded for genome-dependent but biologically incorrect answers.

\textbf{GRPO advantage normalization.}
After rewards are gathered across distributed workers, completions are grouped by their original prompt. For a prompt group of size \(G=\texttt{num\_generations}\), the sequence-level advantage for completion \(j\) is:
\[
A_j =
\frac{R_j - \mu_G}{\sigma_G + 10^{-4}},
\]
where \(\mu_G\) and \(\sigma_G\) are the mean and standard deviation of the rewards within the same prompt group. The final token-level advantage is:
\[
A_{j,i}^{\mathrm{token}}
=
A_j\mathbb{1}[i\in \mathrm{answer}]
+
w_{\mathrm{tool}}R_{\mathrm{tool},i}
+
w_{\mathrm{gene}}R_{\mathrm{gene},i}
+
w_{\mathrm{attn}}R_{\mathrm{attn},i}.
\]
In the main counterfactual launcher, \texttt{attention\_reward\_weight}=0.0, so the attention-based shaping term is disabled.

\textbf{Policy loss.}
The trainer computes current token log-probabilities under the genome-aware policy and compares them with the old log-probabilities from the sampling policy. For each optimized token,
\[
\rho_i =
\exp\left(
\log \pi_\theta(y_i)-\log \pi_{\mathrm{old}}(y_i)
\right).
\]
The clipped GRPO token loss is:
\[
\mathcal{L}_i
=
-\min\left(
\rho_i A_i,\;
\mathrm{clip}(\rho_i,1-\epsilon,1+\epsilon)A_i
\right),
\]
with \(\epsilon=0.2\). The final loss averages over valid model-generated tokens:
\[
\mathcal{L}
=
\frac{1}{B}\sum_{b=1}^{B}
\frac{\sum_i m_{b,i}\mathcal{L}_{b,i}}{\sum_i m_{b,i}},
\]
where \(m_{b,i}\) masks padding tokens and environment-inserted tool payloads.

\subsection{Implementation details}
\label{app:hyperpram}
We run all training experiments on a single 8-GPU node equipped with NVIDIA H200 GPUs under Ubuntu 22.04. The training pipeline consists of three stages: genome--text alignment, agentic supervised fine-tuning, and agentic reinforcement learning. Unless otherwise specified, all stages use full-parameter fine-tuning with bf16 precision. We implement agentic SFT with LLaMA-Factory~\cite{zheng2024llamafactory}, and build the agentic RL training framework on OpenR1~\cite{openr1} and TRL~\cite{vonwerra2020trl}.

First, we conduct gene-text instruction tuning starting from a gene-aligned Qwen3.5-4B checkpoint. The training data consist of gene-text alignment examples, and the goal of this stage is to improve the model's ability to map genome-conditioned representations to textual biological descriptions before introducing tool-use supervision. We train for one epoch with maximum sequence length 1,024, effective batch size 256, and learning rate $2\times10^{-5}$. This stage requires approximately 16 hours on 8 H200 GPUs.

Second, we perform agent-supervised fine-tuning on distilled tool-use trajectories. The trajectories are represented in a multi-turn format containing system prompts, tool specifications, model actions, tool observations, intermediate reasoning, and final structured answers. This stage teaches the model the tool-calling protocol and how to integrate external observations into its final biological prediction. To support long tool-use contexts, we use maximum sequence length 16,384. We train for one epoch with effective batch size 128 and learning rate $5\times10^{-6}$, which takes approximately 8 hours on the same hardware.

Finally, we apply GRPO to improve biological correctness, tool-use behavior, and reliance on the genome modality beyond imitation learning. For each prompt, the policy samples a group of 4 trajectories, and rewards are normalized within the group to obtain relative advantages. Rollouts are generated using colocated vLLM with tensor parallelism over 8 GPUs. We use a global rollout batch size of 64, learning rate $1\times10^{-6}$, maximum completion length 8,192, and a limit of 5 tool-calling iterations per trajectory. Tool-call tokens are optimized as policy actions, while environment-provided tool observations are excluded from the policy loss. During this RL stage, the genome-fusion module remains frozen and only the language-policy parameters are updated. The reward includes terms for JSON validity, biological correctness, scheduled tool use, unsupported direct-answer penalties, and counterfactual gene grounding. The GRPO stage takes approximately 40 hours on 8 H200 GPUs.

\subsection{Evaluation metrics}
We evaluate microbial trait prediction across five groups: interval-valued traits, optimal scalar traits, substrate-source traits, single-label physiological traits, and morphological traits. All model outputs are parsed as JSON and mapped to the corresponding target field. For categorical and substrate-source traits, predictions are further canonicalized using task-specific vocabularies and alias mappings.

For interval-valued traits, including growth temperature, pH, and salinity ranges, we compare the predicted interval $[\hat{l},\hat{u}]$ with the ground-truth interval $[l,u]$. We report interval coverage rate (ICR), which measures whether the predicted interval fully covers the ground-truth interval:
\[
\mathrm{ICR}
=
\frac{1}{N}\sum_{i=1}^{N}
\mathbf{1}(\hat{l}_i\le l_i \land \hat{u}_i\ge u_i).
\]

For optimal scalar traits, including optimal growth temperature, optimal pH, and optimal salinity, predictions are evaluated using root mean squared error (RMSE). If a model outputs an interval-form prediction for an optimal scalar trait, we convert the interval to a scalar using its midpoint before computing RMSE.

For substrate-source traits, including carbon source, electron donor, electron acceptor, and nitrogen source, predictions are treated as ranked label lists. We evaluate the top five predictions using mean average precision at five (mAP@5). Given the ground-truth label set $Y_i$ and ranked predictions $\hat{y}_{i,1},\ldots,\hat{y}_{i,5}$, AP@5 is computed as
\[
\mathrm{AP@5}_i
=
\frac{1}{\min(|Y_i|,5)}
\sum_{k=1}^{5}
\mathrm{Prec@}k_i \cdot \mathbf{1}(\hat{y}_{i,k}\in Y_i),
\]
where $\mathrm{Prec@}k_i$ denotes precision among the top-$k$ predictions. We report mAP@5 by averaging AP@5 across evaluation samples.

For single-label physiological and morphological traits, including oxygen tolerance, energy type, photosynthesis type, carbon fixation pathway, cell shape, motility, flagella, spore formation, and Gram stain, we report accuracy after label canonicalization.

\section{Additional Experimental Results}
\label{app:experimental_details}

% \subsection{Optimization Stability}
% \label{app:optimization_stability}

\subsection{Training Stability Across Stages}
\label{app:training_stability}

\begin{figure}[h]
    \centering
    \includegraphics[width=0.8\linewidth]{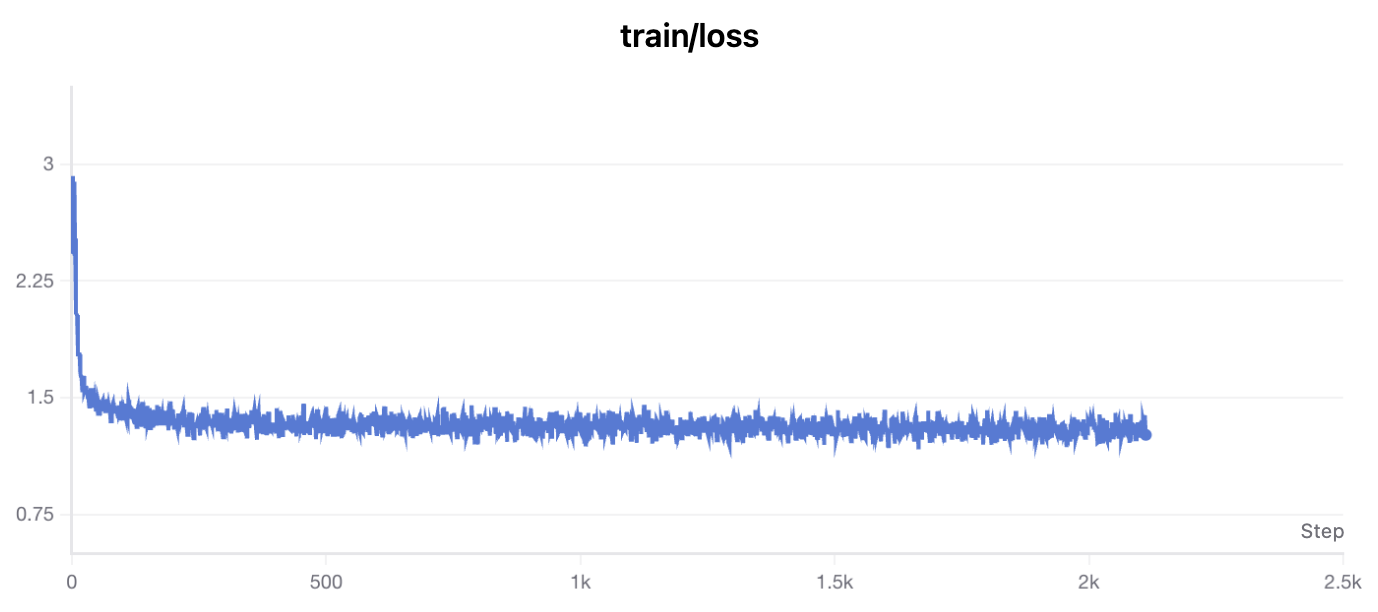}
    \vspace{-0.1cm}
    \caption{Training loss of the first modal-fusion stage, where only the MLP projector is tuned.}
    \label{fig:loss_fusion1}
\end{figure}

\begin{figure}[h]
    \centering
    \includegraphics[width=0.8\linewidth]{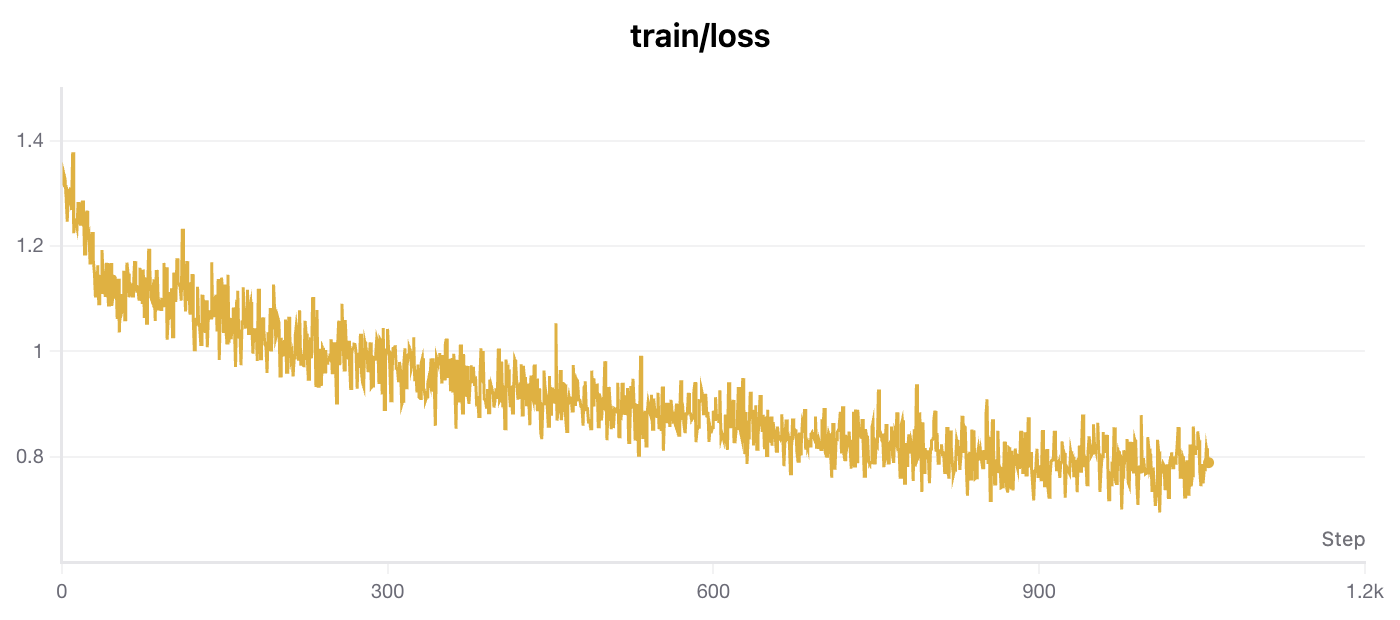}
    \vspace{-0.1cm}
    \caption{Training loss of the second modal-fusion stage, where both the MLP projector and LLM backbone are tuned.}
    \label{fig:loss_fusion2}
\end{figure}

\begin{figure}[h]
    \centering
    \includegraphics[width=0.8\linewidth]{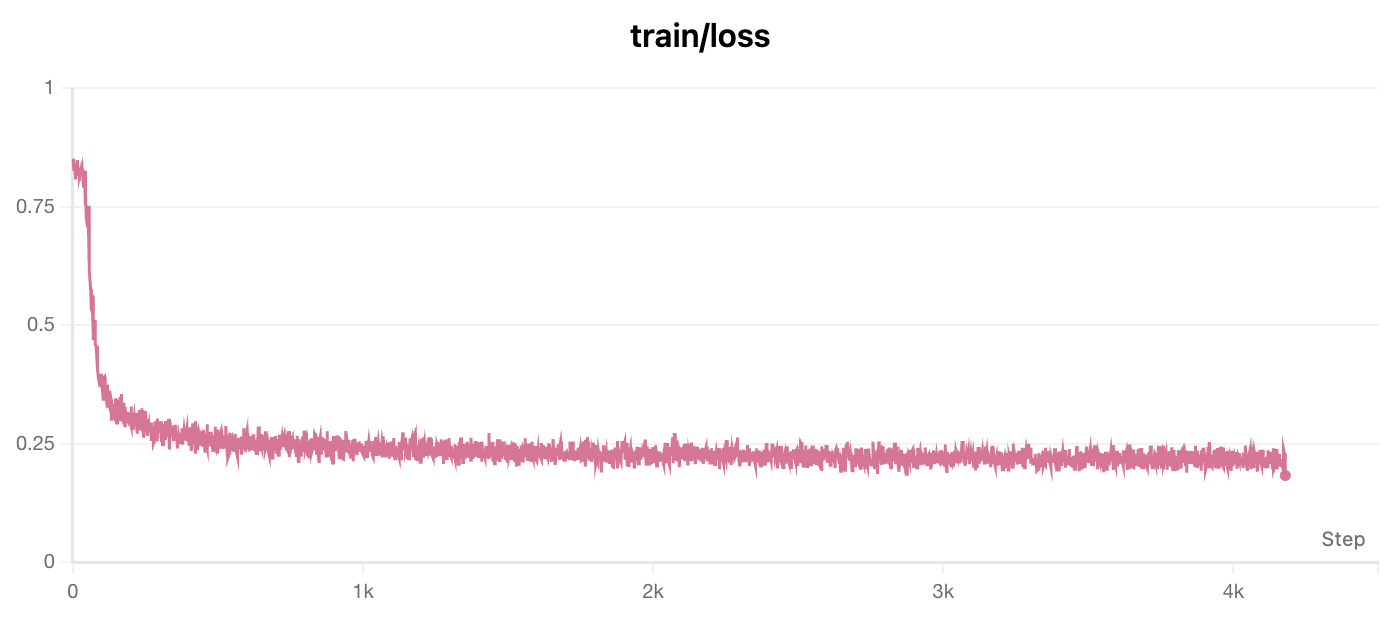}
    \vspace{-0.1cm}
    \caption{Training loss of the agentic SFT stage, where only the LLM policy is tuned on distilled tool-use trajectories.}
    \label{fig:loss_agent_sft}
\end{figure}

\textbf{Training Curves.} 
Figures~\ref{fig:loss_fusion1}--\ref{fig:loss_agent_sft} show the training losses across the two modal-fusion stages and the agentic SFT stage. In the first stage, tuning only the MLP projector yields a steadily decreasing loss, indicating that the projected LucaOne representation can be stably aligned with the LLM token space. In the second stage, jointly tuning the projector and LLM further reduces the loss without visible divergence, suggesting that the model can refine genome--text alignment while preserving stable optimization. The agentic SFT curve also decreases smoothly and stabilizes on long-context tool-use trajectories, confirming that the model can learn the tool-calling format and evidence-conditioned response pattern reliably before GRPO optimization.

\subsection{Frontier LLM Comparison under Strain-Name Input}
\label{app:frontier_llm_strain_name}

Table~\ref{tab:llm_comparison} reports the performance of frontier LLMs when strain names are directly provided. This setting favors general-purpose LLMs because strain names expose taxonomic priors learned during pretraining. The results show that no single frontier model dominates all task groups. GLM-4.7 performs best on physiological-boundary prediction, achieving the highest ICR on growth-temperature, pH, and salinity ranges. For physiological optima, Kimi-K2 performs best on optimal growth temperature and pH, while DeepSeek-V3.2 obtains the lowest RMSE on optimal salinity.

For metabolism and categorical traits, the best model varies by target. GLM-4.7 is strongest on most substrate-source tasks, whereas Kimi-K2 performs best on several categorical and morphology targets, including carbon fixation pathway, energy type, flagella, and Gram stain. These results suggest that strain-name prompting can activate useful biological priors, but the resulting performance is task-dependent and model-dependent. Compared with these large baselines, GGBound remains competitive in the main strain-name analysis, indicating that genome conditioning and biological tool use can compensate for substantially smaller model scale.

\begin{table}[t]
\centering
\caption{
Comparison with frontier LLMs under strain-name input. Best results are shown in \textbf{bold}; second-best results are \underline{underlined}.
}
\label{tab:llm_comparison}
\footnotesize
\setlength{\tabcolsep}{3pt}
\renewcommand{\arraystretch}{0.98}
\begin{tabular}{@{}llccccc@{}}
\toprule
\textbf{\shortstack[l]{Task Group}} 
& \textbf{Prediction Target} 
& \textbf{Metric}
& \textbf{DeepSeek-V3.2} 
& \textbf{DeepSeek-R1} 
& \textbf{GLM-4.7} 
& \textbf{Kimi-K2} \\
\midrule

\multirow{3}{*}{\textbf{\shortstack[l]{Physiological\\Boundary}}}
& Growth temperature range & ICR $\uparrow$
& 0.285 & 0.313 & \textbf{0.420} & \underline{0.339} \\
& pH range & ICR $\uparrow$
& 0.177 & 0.179 & \textbf{0.371} & \underline{0.243} \\
& Salinity range & ICR $\uparrow$
& 0.548 & \underline{0.606} & \textbf{0.760} & 0.555 \\

\midrule
\multirow{3}{*}{\textbf{\shortstack[l]{Physiological\\Optimum}}}
& Optimal growth temperature & RMSE $\downarrow$
& 6.494 & 6.621 & \underline{5.671} & \textbf{5.626} \\
& Optimal pH & RMSE $\downarrow$
& 0.684 & \underline{0.674} & 0.678 & \textbf{0.668} \\
& Optimal salinity & RMSE $\downarrow$
& \textbf{2.088} & 2.315 & \underline{2.276} & 2.283 \\

\midrule
\multirow{4}{*}{\textbf{Metabolism}}
& Carbon source & mAP@5 $\uparrow$
& 0.559 & \textbf{0.590} & \underline{0.562} & 0.522 \\
& Electron acceptor & mAP@5 $\uparrow$
& 0.334 & 0.327 & \textbf{0.407} & \underline{0.351} \\
& Electron donor & mAP@5 $\uparrow$
& \underline{0.605} & 0.603 & \textbf{0.635} & 0.569 \\
& Nitrogen source & mAP@5 $\uparrow$
& 0.175 & 0.139 & \textbf{0.276} & \underline{0.264} \\

\midrule
\multirow{4}{*}{\textbf{\shortstack[l]{Categorical\\Physiology}}}
& Carbon fixation pathway & Accuracy $\uparrow$
& \underline{0.554} & 0.500 & 0.473 & \textbf{0.662} \\
& Energy type & Accuracy $\uparrow$
& 0.776 & \underline{0.784} & 0.709 & \textbf{0.825} \\
& Oxygen tolerance & Accuracy $\uparrow$
& \textbf{0.499} & 0.406 & 0.327 & \underline{0.475} \\
& Photosynthesis type & Accuracy $\uparrow$
& \textbf{0.902} & \underline{0.895} & 0.850 & 0.820 \\

\midrule
\multirow{5}{*}{\textbf{Morphology}}
& Cell shape & Accuracy $\uparrow$
& \underline{0.778} & \textbf{0.782} & 0.764 & \underline{0.778} \\
& Flagella & Accuracy $\uparrow$
& 0.460 & 0.458 & \underline{0.465} & \textbf{0.567} \\
& Gram stain & Accuracy $\uparrow$
& \underline{0.910} & 0.896 & 0.882 & \textbf{0.912} \\
& Motility & Accuracy $\uparrow$
& \textbf{0.772} & \underline{0.756} & 0.689 & 0.752 \\
& Spore formation & Accuracy $\uparrow$
& \underline{0.916} & \textbf{0.922} & 0.911 & \textbf{0.922} \\

\bottomrule
\end{tabular}
\end{table}

\subsection{GGBound Evaluation with Strain-Name Input and Tool Access}
\label{app:strain_name_tool}

\begin{table}[ht]
\centering
\caption{Results with strain-name input and tool access. Best/second-best are bold/underlined.}
\label{tab:appendix_strain_name_tool}
\small
\setlength{\tabcolsep}{3pt}
\footnotesize
\renewcommand{\arraystretch}{0.98}
\begin{tabular}{@{}llccccc@{}}
\toprule
\textbf{\shortstack[l]{Tasks}} 
& \textbf{Field} 
& \textbf{Metric}
& \textbf{Base Model} 
& \textbf{Fusion Model} 
& \textbf{SFT Agent} 
& \textbf{RL Agent} \\
\midrule

\multirow{3}{*}{\textbf{\shortstack[l]{Physiological\\Boundary}}}
& Growth temperature range & ICR $\uparrow$
& 0.172 & 0.208 & \underline{0.298} & \textbf{0.302} \\
& pH range & ICR $\uparrow$
& 0.305 & 0.369 & \underline{0.400} & \textbf{0.407} \\
& Salinity range & ICR $\uparrow$
& \underline{0.606} & 0.604 & \textbf{0.727} & \textbf{0.727} \\

\midrule
\multirow{3}{*}{\textbf{\shortstack[l]{Physiological\\Optimum}}}
& Optimal growth temperature & RMSE $\downarrow$
& 6.599 & 5.923 & \underline{4.413} & \textbf{4.298} \\
& Optimal pH & RMSE $\downarrow$
& 0.886 & 1.135 & \underline{0.665} & \textbf{0.649} \\
& Optimal salinity & RMSE $\downarrow$
& 3.336 & 2.904 & \underline{2.022} & \textbf{2.001} \\

\midrule
\multirow{4}{*}{\textbf{Metabolism}}
& Carbon source & mAP@5 $\uparrow$
& \textbf{0.677} & 0.237 & 0.469 & \underline{0.557} \\
& Electron acceptor & mAP@5 $\uparrow$
& \textbf{0.583} & 0.456 & 0.487 & \underline{0.498} \\
& Electron donor & mAP@5 $\uparrow$
& \textbf{0.748} & 0.258 & 0.649 & \underline{0.652} \\
& Nitrogen source & mAP@5 $\uparrow$
& \textbf{0.446} & 0.334 & 0.420 & \underline{0.425} \\

\midrule
\multirow{4}{*}{\textbf{\shortstack[l]{Categorical\\Physiology}}}
& Carbon fixation pathway & Accuracy $\uparrow$
& \underline{0.694} & 0.516 & 0.661 & \textbf{0.710} \\
& Energy type & Accuracy $\uparrow$
& 0.860 & 0.710 & \underline{0.879} & \textbf{0.888} \\
& Oxygen tolerance & Accuracy $\uparrow$
& 0.713 & \underline{0.732} & 0.726 & \textbf{0.734} \\
& Photosynthesis type & Accuracy $\uparrow$
& 0.842 & \underline{0.863} & \textbf{0.926} & \textbf{0.926} \\

\midrule
\multirow{5}{*}{\textbf{Morphology}}
& Cell shape & Accuracy $\uparrow$
& 0.791 & 0.771 & \underline{0.810} & \textbf{0.813} \\
& Flagella & Accuracy $\uparrow$
& 0.725 & 0.544 & \underline{0.751} & \textbf{0.768} \\
& Gram stain & Accuracy $\uparrow$
& 0.906 & 0.911 & \underline{0.947} & \textbf{0.950} \\
& Motility & Accuracy $\uparrow$
& 0.797 & 0.800 & \underline{0.828} & \textbf{0.837} \\
& Spore formation & Accuracy $\uparrow$
& \underline{0.933} & 0.904 & \textbf{0.957} & \textbf{0.957} \\

\bottomrule
\end{tabular}
\end{table}

Table~\ref{tab:appendix_strain_name_tool} evaluates the GGBound model family when strain names and tool access are both available. Compared with the anonymized benchmark, this setting provides additional taxonomic cues and therefore serves as a complementary upper-bound-style evaluation. Agentic training improves most continuous and categorical targets: the RL agent achieves the best or tied-best performance on all physiological-boundary and optimum tasks, reducing optimal growth-temperature RMSE from 6.599 to 4.298 and optimal salinity RMSE from 3.336 to 2.001 relative to the base model.

The metabolism results show a different trend. The base model performs best on all four substrate-source tasks, suggesting that strain names already provide strong metabolic priors for Qwen3.5-4B. Direct modal fusion can even degrade performance on several metabolism targets, indicating possible conflict between injected genome embeddings and name-based priors before agentic training. In contrast, SFT and RL recover stronger performance on categorical physiology and morphology, with the RL agent achieving the best or tied-best results on most of these targets. Overall, these results show that strain names provide useful prior knowledge, while tool-conditioned training helps integrate such priors with external observations and genome-conditioned evidence.

\subsection{Licenses and Third-Party Resource Usage}
\label{app:licenses}

We use third-party data resources, pretrained models, and open-source software under their respective licenses and terms of use. Table~\ref{tab:licenses} summarizes the main external resources used in this work and their roles in our pipeline.

\begin{table}[t]
\centering
\caption{Third-party resources and licenses. We list the main data sources, pretrained models, and software frameworks used for benchmark construction, model training, tool execution, and baseline evaluation.}
\label{tab:licenses}
\footnotesize
\setlength{\tabcolsep}{3pt}
\begin{tabular}{@{}p{0.24\linewidth}p{0.43\linewidth}p{0.25\linewidth}@{}}
\toprule
\textbf{Resource} & \textbf{Usage in this work} & \textbf{License / terms} \\
\midrule
IJSEM articles & Source literature for strain-level evidence extraction and physiological trait curation. & Publisher copyright and platform terms; used for scholarly research extraction. \\
NCBI genome and protein records & Genome/protein matching and construction of strain-level genome representations. & NCBI molecular data usage policy; NCBI itself places no reuse restrictions on molecular data. \\
BacDive & Structured strain phenotype annotations for alignment, benchmark construction, and agent training. & Creative Commons Attribution 4.0 International License. \\
LucaOne & Frozen genome encoder used to produce gene-level and strain-level embeddings. & Apache License 2.0. \\
Qwen3.5 & LLM backbone for GGBound and teacher model for trajectory generation. & Apache License 2.0. \\
DeepSeek-V3.2 / DeepSeek-R1 & LLM-based extraction, quality scoring, and frontier-model comparison. & MIT License. \\
GLM-4.7 & Frontier-model baseline under strain-name evaluation. & MIT License. \\
Kimi-K2 & Frontier-model baseline under strain-name evaluation. & Modified MIT License. \\
CarveMe & Reconstruction of strain-specific genome-scale metabolic models. & Apache License 2.0. \\
BiGG / iAF692 resources & Source of the archaeal biomass template used in GEM construction. & BiGG/MetaNetX-linked resources under attribution-based terms. \\
LLaMA-Factory & Agentic supervised fine-tuning implementation. & Apache License 2.0. \\
OpenR1 & Base framework for agentic reinforcement learning implementation. & Apache License 2.0. \\
TRL & GRPO and reinforcement-learning training utilities. & Apache License 2.0. \\
vLLM & Batched inference and rollout generation. & Apache License 2.0. \\
\bottomrule
\end{tabular}
\end{table}

For literature-derived data, IJSEM articles are used as source evidence for extraction, normalization, and benchmark construction. We do not redistribute the original full-text articles or publisher-owned article content. Instead, the released benchmark contains normalized strain-level fields, anonymized evaluation instances, and derived labels constructed through our evidence-to-label pipeline. NCBI genome and protein records are used for strain matching and genome representation construction. BacDive annotations are used as structured phenotype evidence and are cited as a core data source; we follow its attribution requirements and do not claim ownership over the original BacDive records.

For models and software, we use open-weight models and open-source frameworks under their corresponding permissive licenses. LucaOne provides the frozen genome encoder; Qwen3.5 provides the trainable language backbone and teacher model; DeepSeek, GLM, and Kimi models are used only for extraction assistance or baseline evaluation. CarveMe and BiGG-derived resources are used to construct GEM-based tool observations, while LLaMA-Factory, OpenR1, TRL, and vLLM support supervised fine-tuning, reinforcement learning, and rollout generation. Our released code and derived benchmark artifacts will include license notices and citations for these third-party resources. 

%%%%%%%%%%%%%%%%%%%%%%%%%%%%%%%%%%%%%%%%%%%%%%%%%%%%%%%%%%%%

% \clearpage
% \input{checklist.tex}

\end{document}